\documentclass[a4paper,UKenglish]{lipics-v2016-arxiv}

\usepackage{microtype}

\usepackage[nocompress]{cite}
\addto\extrasUKenglish{%
}

\usepackage{xspace}
\usepackage{booktabs}

\newenvironment{code}{\noindent%
\begin{tabbing}%
\hspace{2em}\=\hspace{2em}\=\hspace{2em}\=\hspace{2em}\=\hspace{2em}\=%
\hspace{2em}\=\hspace{2em}\=\hspace{2em}\=\hspace{2em}\=\hspace{2em}\=%
\kill}{\end{tabbing}}
\newcommand{\RRem}[1]   {\`{\bf --\hspace{0.5mm}--~}{\rm#1}}

\newcommand{\Oh}[1]{\mathcal{O}\!\left( #1\right)}
\newcommand{\Th}[1]{\Theta\!\left( #1\right)}
\newcommand{\Om}[1]{\Omega\!\left( #1\right)}

\newcommand{\VarArray}{A\xspace}
\newcommand{\BaseCaseSize}{n_0\xspace}
\newcommand{\BlockSize}{b\xspace}
\newcommand{\OversamplingFactor}{\alpha\xspace}

\newcommand{\VarBucketCount}{k\xspace}
\newcommand*{\VarBucket}[1][]{\ifthenelse{\equal{#1}{}}{b}{b_{#1}}\xspace}
\newcommand{\TargetBucketIndex}{\text{dest}}

\newcommand*{\writeblock}[1][]{\ifthenelse{\equal{#1}{}}{w}{w_{#1}}\xspace}
\newcommand*{\readblock}[1][]{\ifthenelse{\equal{#1}{}}{r}{r_{#1}}\xspace}
\newcommand*{\delimiterblock}[1][]{\ifthenelse{\equal{#1}{}}{d}{d_{#1}}\xspace}

\newcommand{\VarThreadCount}{t\xspace}
\newcommand{\OverpartitionFactor}{\beta\xspace}

\newcommand{\algoissssort}{\textsf{IS$^4$o}\xspace}
\newcommand{\algossssort}{\textsf{s$^3$-sort}\xspace}
\newcommand{\algoiparassssort}{\textsf{IPS$^4$o}\xspace}

\newcommand{\algopsort}{\textsf{MCSTLmwm}\xspace}
\newcommand{\algopbalancedsort}{\textsf{MCSTLbq}\xspace}
\newcommand{\algopunbalancedsort}{\textsf{MCSTLubq}\xspace}
\newcommand{\algoppbbs}{\textsf{PBBS}\xspace}
\newcommand{\algoptbb}{\textsf{TBB}\xspace}

\newcommand{\algosblock}{\textsf{BlockQ}\xspace}
\newcommand{\algossort}{\textsf{std-sort}\xspace}
\newcommand{\algosyaros}{\textsf{DualPivot}\xspace}

\newcommand{\pcintelfour}{\mbox{Intel4S}\xspace}
\newcommand{\pcinteltwo}{\mbox{Intel2S}\xspace}
\newcommand{\pcamd}{\mbox{AMD1S}\xspace}

\newcommand{\distuniform}{Uniform\xspace}
\newcommand{\distexpo}{Exponential\xspace}
\newcommand{\distalmostsorted}{AlmostSorted\xspace}
\newcommand{\distsorted}{Sorted\xspace}
\newcommand{\distreversesorted}{ReverseSorted\xspace}
\newcommand{\distones}{Ones\xspace}
\newcommand{\distduplicatesroot}{RootDup\xspace}
\newcommand{\distduplicatestwice}{TwoDup\xspace}
\newcommand{\distduplicateseight}{EightDup\xspace}

\newcommand{\bytes}{100Bytes\xspace}
\newcommand{\pair}{Pair\xspace}
\newcommand{\quartet}{Quartet\xspace}

\title{In-place Parallel Super Scalar Samplesort (\algoiparassssort)}

\author{Michael Axtmann}
\author{Sascha Witt}
\author{Daniel Ferizovic}
\author{Peter Sanders}
\affil{Karlsruhe Institute of Technology, Karlsruhe, Germany\\
  \texttt{\{michael.axtmann,sascha.witt,sanders\}@kit.edu}}
\authorrunning{M.\ Axtmann, S.\ Witt, D.\ Ferizovic, P.\ Sanders}

\Copyright{Michael Axtmann, Sascha Witt, Daniel Ferizovic, Peter Sanders}

\DOIPrefix{}

\begin{document}

\maketitle

\begin{abstract}
  We present a sorting algorithm that works in-place, executes in
  parallel, is cache-efficient, avoids branch-mispredictions, and
  performs work $\Oh{n\log n}$ for arbitrary inputs with high
  probability.  The main algorithmic contributions are new ways to
  make distribution-based algorithms in-place: On the practical side,
  by using coarse-grained block-based permutations, and on the theoretical
  side, we show how to eliminate the recursion stack.  Extensive
  experiments show that our algorithm \algoiparassssort\ scales well
  on a variety of multi-core machines. We outperform our closest
  in-place competitor by a factor of up to 3.  Even as a sequential
  algorithm, we are up to 1.5 times faster than the closest sequential
  competitor, BlockQuicksort.
\end{abstract}

\section{Introduction}

Sorting an array $\VarArray[1..n]$ of $n$ elements according to a total ordering of their keys is a
fundamental subroutine used in many applications.  Sorting is used for
index construction, for bringing similar elements
together, or for processing data in a ``clever'' order.
Indeed, often sorting is the most expensive part of a program.
Consequently, a huge amount of research on sorting has been done.
In particular, algorithm engineering has studied how to make sorting practically fast
in presence of complex features of modern hardware like
multi-core (e.g.,~\cite{TsiZha03,putze2007mcstl,blelloch2010low,shun2012brief}),
instruction parallelism (e.g.,~\cite{sanders2004super}),
branch prediction  (e.g.,~\cite{sanders2004super,kaligosi2006branch,KS06,edelkamp2016blockquicksort}),
caches (e.g.,~\cite{sanders2004super,BFV04,Fran04,blelloch2010low}),
or virtual memory (e.g.,~\cite{Rah03,jurkiewicz2015model}).  In
contrast, the sorting algorithms used in the standard libraries of
programming languages like Java or C++ still use variants of quicksort
-- an algorithm that is more than 50 years old.  A reason seems to be
that you have to outperform quicksort in every respect in order to
replace it.  This is less easy than it sounds since quicksort is a
pretty good algorithm -- it needs $\Oh{n\log n}$ expected work, it can
be parallelized~\cite{TsiZha03,putze2007mcstl}, it can be implemented
to avoid branch mispredictions~\cite{edelkamp2016blockquicksort}, and
it is reasonably cache-efficient.  Perhaps most importantly, quicksort
works (almost) in-place%
\footnote{In algorithm theory, an algorithm works in-place if it uses only constant space in
  addition to its input. We use the term \emph{strictly in-place} for this case.
  In algorithm engineering, one is sometimes satisfied if the additional space is sublinear
  in the input size. We adopt this convention but use the term \emph{almost in-place}
  when we want to make clear what we mean. Quicksort needs logarithmic additional space.}
which is of crucial importance for very large inputs. This feature rules out many contenders.
Further algorithms are eliminated by the requirement to work for arbitrary data types and input distributions.
This makes integer sorting algorithms like radix sort (e.g.,~\cite{kokot2017even}) or using specialized hardware (e.g.,~GPUs or SIMD instructions) less attractive,
since these algorithms cannot be used in a reusable library where they have to work for arbitrary data types.
Another portability issue is that the algorithm should use no code specific to the processor architecture or the operating system like non-temporal writes or overallocation of virtual memory (e.g. \cite{SanWas11}).
One aspect of making an algorithm in-place is that such ``tricks'' are not needed.
Hence, this paper focuses on portable comparison-based algorithms and also considers how the algorithms can be made robust for arbitrary inputs, e.g., with a large number of repeated keys.

The main contribution of this paper is to propose a new algorithm -- \emph{{\bf\em I}n-place {\bf\em P}arallel {\bf\em S}uper {\bf\em S}calar {\bf\em S}ample{\bf\em so}rt}~(\algoiparassssort)%
\footnote{The Latin word ``ipso'' means ``by itself'', referring to the in-place feature of \algoiparassssort.}
-- that combines enough advantages to become an attractive replacement of quicksort.
Our starting point is \emph{super scalar samplesort}~(\algossssort)~\cite{sanders2004super} which already provides a very good sequential non-in-place algorithm that is cache-efficient,
  allows considerable instruction parallelism, and avoids branch mispredictions.
\algossssort\ is a variant of samplesort, which in turn is a generalization of quicksort to multiple pivots.
The main operation is distributing elements of an input sequence to $\VarBucketCount$ output buckets of about equal size.
We parallelize this algorithm using $\VarThreadCount$ threads and make it more robust by taking advantage of inputs with many identical keys.
Our main innovation is to make the algorithm in-place. The first phase of \algoiparassssort\ distributes the elements to $\VarBucketCount$ \emph{buffer} blocks.
When a buffer becomes full, it is emptied into a block of the input array that has already been distributed.
Subsequently, the memory blocks are permuted into the globally correct order.
A cleanup step handles empty blocks and half-filled buffer blocks.
The distribution phase is parallelized by assigning disjoint pieces of the input array to different threads.
The block permutation phase is parallelized using atomic fetch-and-add operations for each block move.
Once subproblems are small enough, they can be solved independently in parallel.

After discussing related work in \autoref{s:related} and
introducing basic tools in \autoref{sec:preliminaries},
we describe our new algorithm
\algoiparassssort\ in \autoref{sec: inplace algo}.
\autoref{s:experiments} makes an experimental evaluation.
An overall discussion and possible future work is given in \autoref{s:conclusion}.
The appendix gives further experimental data and proofs.

\section{Related Work}\label{s:related}

Variants of Hoare's quicksort~\cite{hoare1962quicksort,musser1997introspective} are generally considered some of the most efficient general purpose sorting algorithms.
Quicksort works by selecting a \emph{pivot} element and partitioning the array such that all elements smaller than the pivot are in the left part and all elements larger than the pivot are in the right part.
The subproblems are solved recursively.
A variant of quicksort (with a fallback to heapsort to avoid worst case scenarios) is currently used in the C++ standard library of GCC~\cite{musser1997introspective}.
Some variants of quicksort use two or three pivots~\cite{yaroslavskiy2009dual,KLMQ14} and achieve improvements of around 20\% in running time over the single-pivot case.
Dual-pivot quicksort~\cite{yaroslavskiy2009dual} is the default sorting routine in Oracle~Java~7 and~8.
The basic principle of quicksort remains, but elements are partitioned into three or four subproblems instead of two.
Increasing the number of subproblems (from now on called \emph{buckets}) even further leads to samplesort~\cite{blelloch1991comparison,blelloch2010low}.
Unlike single- and dual-pivot quicksort, samplesort is usually not in-place, but it is well-suited for parallelization and more cache-efficient.

Super~scalar~samplesort~\cite{sanders2004super}~(\algossssort) improves on samplesort by avoiding inherently hard-to-predict conditional branches linked to element comparisons.
Branch mispredictions are very expensive
because they disrupt the pipelined and instruction-parallel operation of modern processors.
Traditional quicksort variants suffer massively from branch mispredictions~\cite{kaligosi2006branch}.
By replacing conditional branches with conditionally executed machine instructions, branch mispredictions can be largely avoided.
This is done automatically by modern compilers if only a few instructions depend on a condition.
As a result, \algossssort is up to two times faster than quicksort~(\texttt{std::sort}), at the cost of $\Oh{n}$ additional space.
BlockQuicksort~\cite{edelkamp2016blockquicksort} applies similar ideas to single-pivot quicksort, resulting in a very fast in-place sorting algorithm.

Super scalar samplesort has also been adapted for efficient parallel string sorting~\cite{bingmann2013parallel}.
Our implementation is influenced by that work with respect to parallelization and handling equal keys.
Moreover, we were also influenced by an implementation of \algossssort\ written by Lorenz Hübschle-Schneider.
A prototypical implementation of sequential non-blocked in-place \algossssort\ in a student project by our student Florian Weber motivated us to develop \algoiparassssort.

The best practical comparison-based multi-core sorting algorithms we have found are based on
multi-way mergesort~\cite{putze2007mcstl} and samplesort~\cite{shun2012brief}, respectively.
The former algorithm is used in the parallel mode of the C++ standard library of GCC.
Parallel in-place algorithms are based on quicksort so far.
Intel's Thread Building Blocks library~\cite{reinders2007intel}
contains a variant that uses only sequential partitioning.
The MCSTL library~\cite{putze2007mcstl} contains two implementations of the more scalable
parallel quicksort by Tsigas and Zhang~\cite{TsiZha03}.

There is a considerable amount of work by the theory community on (strictly) in-place sorting (e.g.,~\cite{Fran04,FraGef05}).
However, there are few -- mostly negative -- results on transferring these results into practice.
Katajainen and Teuhola~\cite{katajainen1996practical} report that in-place mergesort is slower than heapsort,
which is quite slow for big inputs due to its cache-inefficiency.
Chen~\cite{CHEN200634} reports that in-place merging takes about six times longer than non-in-place merging.
There is previous work on (almost) in-place multi-way merging or data distribution. However,
few of these papers seem to address parallelism. There are also other problems. For example,
the multi-way merger in~\cite{Geffert2009} needs to allocate very large blocks to become efficient.
In contrast, the block size of \algoiparassssort\ does not depend on the input size.
In-place data distribution, e.g., for radix sort~\cite{cho2015paradis}, is often done element by element.
Using this for samplesort would require doing the expensive element classification twice and would also make parallelization difficult.

\section{Preliminaries}\label{sec:preliminaries}

\subparagraph*{(Super Scalar) Samplesort.}

Samplesort~\cite{FraMck70} can be viewed as a generalization of quicksort which uses multiple pivots to split the input into $\VarBucketCount$ \emph{buckets} of about equal size.
A robust way for determining the pivots is to sort $\OversamplingFactor \VarBucketCount - 1$~randomly sampled input elements.
The pivots~$s_1$,\ldots $s_{\VarBucketCount-1}$ are then picked equidistantly from the sorted sample.
Element $e$ goes to bucket $\VarBucket[i]$ if~$s_{i-1}\leq e<s_i$ (with $s_0=-\infty$ and $s_\VarBucketCount=\infty$).
The main contribution of \algossssort~\cite{sanders2004super} is to eliminate branch mispredictions for element classification.
Assuming $\VarBucketCount$ is a power of two, the pivots are stored in an array $a$ representing a complete binary search tree:
  $a_1=s_{\VarBucketCount/2}$, $a_2=s_{\VarBucketCount/4}$, $a_3=s_{3\VarBucketCount/4},\ldots$
More generally, the left successor of $a_i$ is $a_{2i}$ and its right successor is $a_{2i+1}$.
Thus, navigating this tree is possible by performing a conditional instruction for incrementing an array index.
We adopt (and refine) this approach to element classification but change the organization of buckets in order to make the algorithm in-place.

\section{In-Place Parallel Super Scalar Samplesort (\algoiparassssort)}\label{sec: inplace algo}

\algoiparassssort is based on the ideas of \algossssort.
It is a recursive algorithm, where each step divides the input into $\VarBucketCount$~buckets,
such that each element of bucket~$\VarBucket[i]$ is smaller than all elements of $\VarBucket[i+1]$.
As long as problems with at least $\OverpartitionFactor \frac{n}{\VarThreadCount}$~elements exist,
  we partition those problems one after another with $\VarThreadCount$~threads in parallel.
Here, $\OverpartitionFactor$ is a tuning parameter.
Then we assign remaining problems in a balanced way to threads, which sort them sequentially.

The partitioning consists of four phases.
{\bf Sampling} determines the bucket boundaries.
{\bf Local classification} groups the input into blocks such that all elements in each block belong to the same bucket.
{\bf Block permutation} brings the blocks into the globally correct order.
Finally, we perform some {\bf cleanup} around the bucket boundaries.
The following sections will explain each of these phases in more detail.

\subparagraph*{Sampling.} The sampling phase is similar to the sampling in \algossssort.
The main difference is that we swap the sample to the front of the input array
to keep the in-place property even if the oversampling factor $\OversamplingFactor$ depends on $n$.

\subsection{Local Classification}

\begin{figure}[tbp]
  \begin{center}
    \includegraphics[]{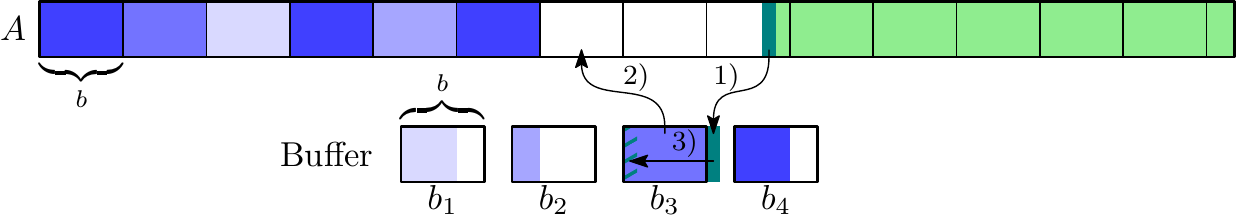}
  \end{center}
  \caption{\label{fig:block gen progress}
  Local classification.
  Blue elements have already been classified, with different shades indicating different buckets.
  Unprocessed elements are green.
  Here, the next element (in dark green) has been determined to belong to bucket~$\VarBucket[3]$.
  As that buffer block is already full, we first write it into the array~$\VarArray$, then write the new element into the now empty buffer.
  }
\end{figure}

\begin{figure}[tbp]
  \begin{center}
  \includegraphics[]{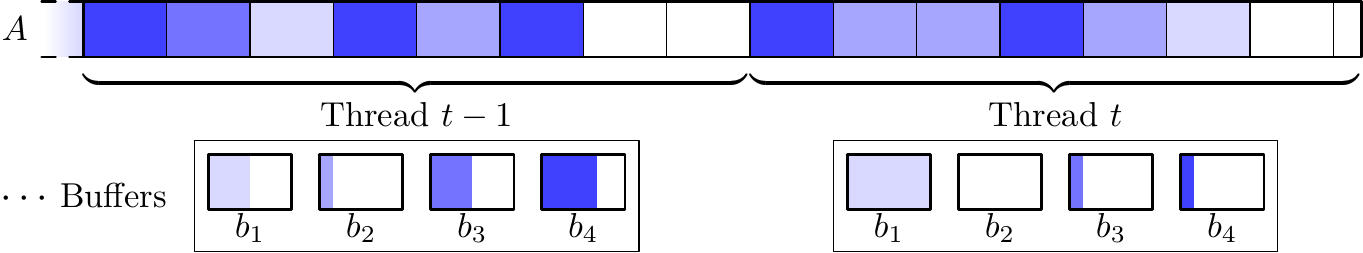}
  \end{center}
  \caption{\label{fig:block generation final}
  Input array and block buffers of the last two threads after local classification.
  }
\end{figure}

The input array $\VarArray$ is viewed as an array of blocks each containing $\BlockSize$~elements (except possibly for the last one).
For parallel processing, we divide the blocks of $\VarArray$ into $\VarThreadCount$~stripes of equal size -- one for each thread.
Each thread works with a local array of $\VarBucketCount$ buffer blocks -- one for each bucket.
A thread then scans its stripe.
Using the search tree created in the previous phase, each element in the stripe is classified into one of the $\VarBucketCount$~buckets,
  then moved into the corresponding local buffer block.
If this buffer is already full, it is first written back into the local stripe, starting at the front.
It is clear that there is enough space to write $\BlockSize$~elements into the local stripe,
  since at least $\BlockSize$ more elements have been scanned from the stripe than have been written back -- otherwise no full buffer could exist.

In this way, each thread creates blocks of $\BlockSize$~elements belonging to the same bucket.
\autoref{fig:block gen progress} shows a typical situation during this phase.
To achieve the in-place property, we do not track which bucket each block belongs to.
However, we do keep count of how many elements are classified into each bucket, since we need this information in the following phases.
This information can be obtained almost for free as a side effect of maintaining the buffer blocks.
\autoref{fig:block generation final} depicts the input array after local classification.
Each stripe contains a number of full blocks, followed by a number of empty blocks.
The remaining elements are still contained in the buffer blocks.

\subsection{Block Permutation}

In this phase, the blocks in the input array will be rearranged such that they appear in the correct order.
From the previous phase we know, for each stripe, how many elements belong to each bucket.
We perform a prefix sum operation to compute the exact boundaries of the buckets in the input array.
In general, these will not coincide with the block boundaries.
For the purposes of this phase, we will ignore this:
We mark the beginning of each bucket $\VarBucket[i]$ with a delimiter pointer~$\delimiterblock[i]$, rounded up to the next block.
We similarly mark the end of the last bucket~$\VarBucket[\VarBucketCount]$ with a delimiter pointer~$\delimiterblock[\VarBucketCount+1]$.
Adjusting the boundaries may cause a bucket to ``lose'' up to $\BlockSize - 1$ elements;
  this doesn't affect us, since this phase only deals with full blocks, and any elements not constituting a full block remain in the buffers.
Additionally, if the input size is not a multiple of $\BlockSize$, some of the $\delimiterblock[i]$s may end up outside the bounds of $\VarArray$.
To avoid overflows, we allocate a single empty \emph{overflow block} which the algorithm will use instead of writing to the final (partial) block.

\begin{figure}[tbp]
  \begin{center}
    \includegraphics[]{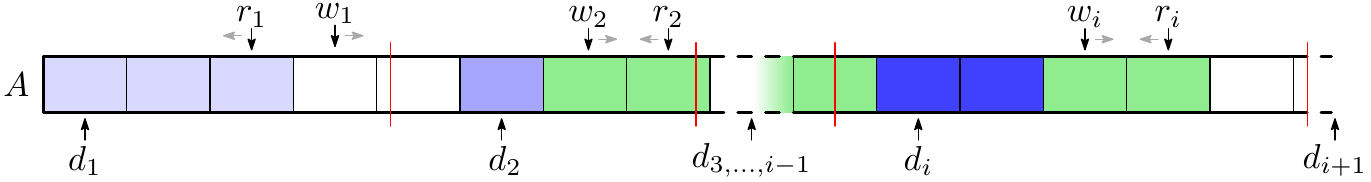}
  \end{center}
  \caption{\label{fig:block perm invariant}
    Invariant during block permutation.
    In each bucket~$\VarBucket[i]$, blocks in $[\delimiterblock[i],\,\writeblock[i])$ are already correct (blue),
      blocks in $[\writeblock[i],\,\readblock[i]]$ are unprocessed (green), and blocks in $[\max(\writeblock[i],\readblock[i]+1),\,\delimiterblock[i+1])$ are empty (white).
  }
\end{figure}

For each $\VarBucket[i]$, a write pointer~$\writeblock[i]$ and a read pointer~$\readblock[i]$ is introduced;
  these will be set such that all unprocessed blocks, i.e., blocks that still need to be moved into the correct bucket,
  are found between $\writeblock[i]$~and~$\readblock[i]$.
During the block permutation, we maintain the following invariant for each bucket $\VarBucket[i]$, visualized in \autoref{fig:block perm invariant}:

\begin{itemize}
  \item Blocks to the left of $\writeblock[i]$ (exclusive) are correctly placed, i.e., contain only elements belonging to $\VarBucket[i]$.
  \item Blocks between $\writeblock[i]$ and $\readblock[i]$ (inclusive) are unprocessed, i.e., may need to be moved.
  \item Blocks to the right of $\max(\writeblock[i],\readblock[i]+1)$ (inclusive) are empty.
\end{itemize}

In other words, each bucket follows the pattern of correct blocks followed by unprocessed blocks followed by empty blocks,
  with $\writeblock[i]$ and $\readblock[i]$ determining the boundaries.
In the parallel case, we may need to establish this invariant by moving some empty blocks to the end of a bucket (see
  \autoref{app:algorithm} for details);
  in the sequential algorithm, the result of the classification phase already has this pattern.
The read pointers~$\readblock[i]$ are then set to the first non-empty block in each bucket, or~$\delimiterblock[i] -1$ if there are none.

We are now ready to start the block permutation.
Each thread maintains two local swap buffers.
We define a \emph{primary} bucket~$\VarBucket[p]$ for each thread;
  whenever both its buffers are empty, a thread tries to read an unprocessed block from its primary bucket.
To do so, it decrements the read pointer~$\readblock[p]$ (atomically) and reads the block it pointed~to into one of its swap buffers.
If $\VarBucket[p]$ contains no more unprocessed blocks (i.e., $\readblock[p] < \writeblock[p]$), it switches its primary bucket to the next bucket (cyclically).
If it completes a whole cycle and arrives back at its initial primary bucket, there are no more unprocessed blocks and this phase ends.
The starting points for the threads are distributed across that cycle to reduce contention.

\begin{figure}[tbp]
  \begin{center}
    \begin{subfigure}[t]{0.496\textwidth}
      \includegraphics[]{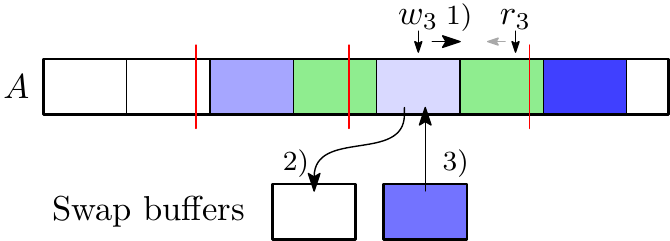}
      \caption{\label{fig:block perm a}\begin{minipage}[t]{0.85\textwidth}
        Swapping a block into its correct position.
      \end{minipage}}
    \end{subfigure}
    \hfill
    \begin{subfigure}[t]{0.496\textwidth}
      \includegraphics[]{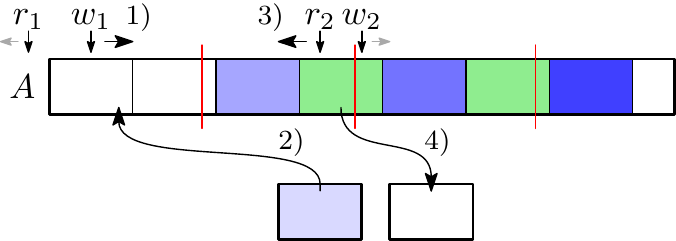}
      \caption{\label{fig:block perm b}\begin{minipage}[t]{0.85\textwidth}
        Moving a block into an empty position, followed by refilling the swap buffer.
      \end{minipage}}
    \end{subfigure}
  \end{center}
  \vspace{-10pt}
  \caption{\label{fig:block perm}
    Block permutation examples.
  }
\end{figure}

Once it has a block, each thread classifies the first element of that block to find its destination bucket~$\VarBucket[\TargetBucketIndex]$.
There are now two possible cases, visualized in \autoref{fig:block perm}:

\begin{itemize}
  \item As long as $\writeblock[\TargetBucketIndex] \leq \readblock[\TargetBucketIndex]$, write pointer $\writeblock[\TargetBucketIndex]$ still points to an unprocessed block in bucket $\VarBucket[\TargetBucketIndex]$. In this case, the thread increases $\writeblock[\TargetBucketIndex]$,
      reads the unprocessed block into its empty swap buffer, and writes the other one into its place.
  \item If $\writeblock[\TargetBucketIndex] > \readblock[\TargetBucketIndex]$, no unprocessed block remains in bucket $\VarBucket[\TargetBucketIndex]$ but $\writeblock[\TargetBucketIndex]$ now points to an empty block. In this case, the thread increases $\writeblock[\TargetBucketIndex]$, writes its swap buffer to the empty block and then reads a new unprocessed block from its primary bucket.
\end{itemize}

We repeat these steps until all blocks are processed.
We can skip unprocessed blocks which are already correctly placed:
We simply classify blocks \emph{before} reading them into a swap buffer, and skip as needed.
We omitted this from the above description for the sake of clarity.
In some cases, this reduces the number of block moves significantly.

It is possible that one thread wants to write to a block that another thread is currently reading from
  (when the reading thread has just decremented the read pointer, but has not yet finished reading the block into its swap buffer).
To avoid data races, we keep track of how many threads are reading from each bucket.
Threads are only allowed to write to empty blocks if no other threads are currently reading from the bucket in question, otherwise they wait.
Note that this situation occurs at most once for each bucket, namely when $\writeblock[\TargetBucketIndex]$ and $\readblock[\TargetBucketIndex]$ cross each other.
In addition, we store each $\writeblock[i]$ and $\readblock[i]$ in a single 128-bit word which we read and modify atomically.
This ensures a consistent view of both pointers for all threads.

\subsection{Cleanup}

After the block permutation, some elements may still be in incorrect positions.
This is due to the fact that we only moved blocks, which may span bucket boundaries.
We call the partial block at the beginning of a bucket its \emph{head} and the partial block at its end its \emph{tail}.

\begin{figure}[tbp]
  \begin{center}
    \includegraphics[]{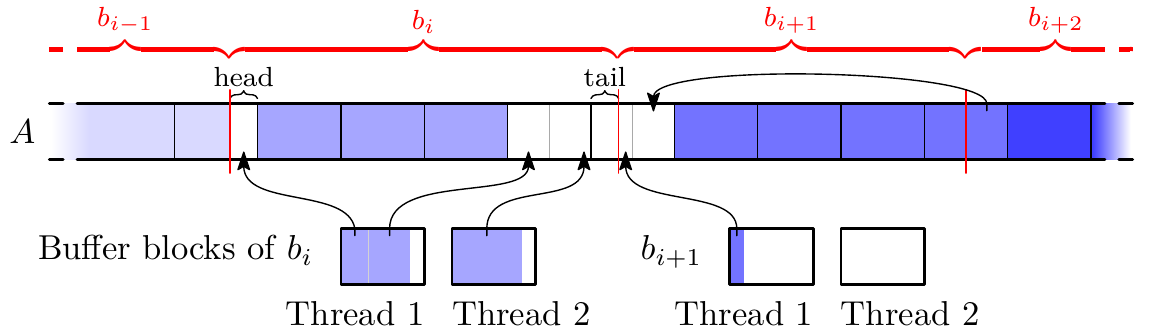}
  \end{center}
  \caption{\label{fig:cleanup}
    An example of the steps performed during cleanup.
  }
\end{figure}

We assign consecutive buckets evenly to threads;
  if $\VarThreadCount > \VarBucketCount$, some threads will not receive any buckets,
  but those that do only need to process a single bucket each.
Each thread reads the head of the first bucket of the next thread into one of its swap buffers.
Then, each thread processes its buckets from left to right, moving incorrectly placed elements into empty array entries.
The incorrectly placed elements of bucket~$\VarBucket[i]$ consist of the elements in the head of $\VarBucket[i+1]$
  (or the swap buffer, for the last bucket), the partially filled buffers from the local classification phase (of all threads),
  and, for the corresponding bucket, the overflow buffer.
Empty array entries consist of the head of $\VarBucket[i]$ and any (empty) blocks to the right of $\writeblock[i]$ (inclusive).
Although the concept is relatively straightforward, the implementation is somewhat involved, due to the many parts that have to be brought together.
\autoref{fig:cleanup} shows an example of the steps performed during this phase.
Afterwards, all elements are back in the input array and correctly partitioned, ready for recursion.

\subsection{The Case of Many Identical Keys}\label{ss:identical}

Having inputs with many identical keys can be a problem for samplesort,
  since this might move large fractions of the keys through many levels of recursion.
We turn such inputs into \emph{easy} instances by introducing separate buckets for elements identical to pivots
  (keys occurring more then $\frac{n}{\VarBucketCount}$ times are likely to become pivots).
Finding out whether an element has to go into an equality bucket (and which one) can be implemented
  using a single additional comparison~\cite{bingmann2013parallel} and, once more, without a conditional branch.
Equality buckets can be skipped during recursion and thus are not a load balancing problem.

\subsection{Analysis}

Algorithm \algoiparassssort\ inherits from \algossssort\ that it has virtually no branch mispredictions (this includes the comparisons for placing elements into equality buckets discussed in \autoref{ss:identical}).
More interesting is the parallel complexity.
Here, the main issue is the number of accesses to main memory.
We analyze this aspect in the parallel external memory (PEM) model~\cite{AGNS08},
  where each of the $t$~threads has a private cache of size~$M$ and access to main memory happens in blocks of size~$B$.
In \autoref{app:analysis}, we prove:
\begin{theorem}\label{thm:io}
  Assuming $b=\Th{tB}$ (buffer block size), $M=\Omega(ktB)$, $n_0=\Oh{M}$ (base case size), $\alpha\in\Omega(\log t)\cap\Oh{t}$ (oversampling factor), and $n=\Om{\max(k,t)t^2B}$, \algoiparassssort\ has an I/O-complexity of
  $\Oh{\frac{n}{tB}\log_k\frac{n}{n_0}}$ block transfers with high probability.
\end{theorem}
Basically, \autoref{thm:io} tells us that \algoiparassssort\ is asymptotically I/O efficient if certain rather steep assumptions on cache size and input size hold.
In particular, the blocks need to have size $b=\Th{tB}$
in order to amortize contention on shared block pointers.
Lifting those could be an interesting theoretical question and we would have to see how absence of branch mispredictions and the in-place property can be combined with previous techniques~\cite{AGNS08,blelloch2010low}.
However, it is likely that the constant factors involved are much larger than for our simple implementation.
Thus, the constant factors will be the main issue in bringing theory and practice further together.
To throw some light on this aspect, let us compare the constant factors in I/O-volume (i.e., data flow between cache and main memory) for  the sequential algorithms \algoissssort~(\algoiparassssort\ with $t=1$) and
\algossssort. To simplify the discussion, we assume a single level of recursion, $k=256$ and $8$-byte elements.
In \autoref{app:analysis}, we show that \algoissssort\ needs about $48n$ bytes of I/O volume, whereas
\algossssort needs (more than) $86n$ -- almost twice that of  \algoissssort.
This is surprising since on first glance, the partitioning algorithm of \algoissssort\ writes the data twice, whereas \algossssort does this only once.
However, this is more than offset by ``hidden'' overheads of \algossssort like memory management, allocation misses, and associativity misses.

Finally, we consider the memory overhead of \algoiparassssort.
In \autoref{app:analysis}, we show:
\begin{theorem}\label{thm:space}
  \algoiparassssort\ requires additional space $\Oh{kbt+\log_k \frac{n}{n_0}}$.
\end{theorem}
In practice, the term $\Oh{kbt}$ (mostly for the distribution buffers) will dominate. However,
for a strictly in-place algorithm in the sense of algorithm theory, we need to get rid of the $\Oh{\log n}$
term which depends on the input size. We discuss this separately in \autoref{ss:strictly}.

\subsection{From Almost In-Place to Strictly In-Place}\label{ss:strictly}

We now explain how the space consumption of \algoiparassssort can be made independent of $n$ in a rather simple way.
We can restrict ourselves to the sequential case, since only $\Oh{\log_kt}$ levels of parallel recursion are needed to arrive at subproblems that are solved sequentially.
We require the partitioning operation to mark the beginning of each bucket by storing the largest element of a bucket in its first entry.
By searching the next larger element, we can then find the end of the bucket.
Note that this is possible in time logarithmic in the bucket size using exponential/binary search.
We assume that the corresponding function $\mathit{searchNextLargest}$ returns $n+1$ if no larger elements exists -- this happens for the last bucket.
The following pseudocode uses this approach to emulate recursion in constant space for sequential \algoissssort.
\begin{code}
  $i := 1$\RRem{first element of current bucket}\\
  $j := n+1$\RRem{first element of next bucket}\\
  \textbf{while} $i<n$ \textbf{do}\+\\
  \textbf{if} $j-i<n_0$ \textbf{then} $\mathit{smallSort}(a,i,j-1)$;\quad $i := j$\RRem{base case}\\
    \textbf{else} $\mathit{partition}(a,i,j-1)$\RRem{partition first unsorted bucket}\\
    $j := \mathit{searchNextLargest}(A[i],A,i+1,n)$\RRem{find beginning of next bucket}
\end{code}

\subsection{Implementation Details}\label{s:details}

The strategy for handling identical keys described in \autoref{ss:identical} is enabled conditionally:
After the splitters have been selected from the initial sample, we check for and remove duplicates.
Equality buckets are only used if there were duplicate splitters.

For buckets under a certain base case size $\BaseCaseSize$, we stop the recursion and fall back on insertion sort.
Additionally, we use an adaptive number of buckets on the last two levels of the recursion,
  such that the expected size of the final buckets remains reasonable.
For example, instead of performing two $256$-way partitioning steps to get $2^{16}$~buckets of 2 elements,
  we might perform two $64$-way partitioning steps to get $2^{12}$~buckets of about $32$~elements.
Furthermore, on the last level, we perform the base case sorting immediately after the bucket
  has been completely filled in the cleanup phase, before processing the other buckets.
This is more cache-friendly, as it eliminates the need for another pass over the data.

\algoiparassssort has several parameters that can be used for tuning and adaptation.
We performed our experiments using (up to) $\VarBucketCount=256$ buckets,
  an oversampling factor of $\OversamplingFactor=0.2 \log n$,
  an overpartitioning factor of $\OverpartitionFactor=1$,
  a base case size of $\BaseCaseSize=16$ elements,
  and a block size of about 2\;KiB, or $\BlockSize=\max\left(1, 2^{\lfloor11 - \log_2 s\rfloor}\right)$ elements,
  where $s$ is the size of an element in bytes.
In the sequential case, we avoid the use of atomic operations on pointers.
All algorithms are written in C++ and compiled with version 6.2.0 of the GNU compiler collection,
  using the optimization flags ``\texttt{-march=native -O3}''.
For parallelization, we employ OpenMP.
Our implementation can be found at \url{https://github.com/SaschaWitt/ips4o}.

\section{Experimental Results}\label{s:experiments}
We present the results of our in-place parallel sorting algorithm \algoiparassssort.
We compare the results of \algoiparassssort with its in-place competitors,
  parallel sort from the Intel\textregistered~TBB library~\cite{reinders2007intel}~(\algoptbb),
  parallel unbalanced quicksort from the GCC STL~library~(\algopunbalancedsort),
  and parallel balanced quicksort from the GCC STL~library~(\algopbalancedsort).
We also give results on the parallel non-in-place sorting algorithms,
  parallel samplesort from the problem based benchmark suite~\cite{shun2012brief}~(\algoppbbs)
  and parallel multiway mergesort from the GCC STL~library~\cite{putze2007mcstl}~(\algopsort).
We also ran sequential experiments and present the results of \algoissssort, the sequential implementation of \algoiparassssort.
We compare the results of \algoissssort with its sequential competitors, a recent implementation~\cite{Lorenz2016ssss}
  of non-in-place Super Scalar Samplesort~\cite{sanders2004super}~(\algossssort) optimized for modern hardware,
  BlockQuicksort~\cite{edelkamp2016blockquicksort}~(\algosblock),
  Dual-Pivot Quicksort~\cite{yaroslavskiy2009dual}~(\algosyaros),
  and introsort from the GCC STL~library~(\algossort).

We ran benchmarks with nine input distributions: Uniformly distributed~(\emph{\distuniform}), exponentially distributed~(\emph{\distexpo}),
  and almost sorted~(\emph{\distalmostsorted}), proposed by Shun~et.~al.~\cite{shun2012brief};
  \emph{\distduplicatesroot}, \emph{\distduplicatestwice},
  and \emph{\distduplicateseight} from Edelkamp~et.~al.~\cite{edelkamp2016blockquicksort}; and \emph{\distsorted} (sorted \distuniform input), \emph{\distreversesorted}, and \emph{\distones} (just ones).
The input distribution \distduplicatesroot sets $\VarArray[i] = i \mod \lfloor\sqrt{n}\rfloor$,
  \distduplicatestwice sets $\VarArray[i] = i^2 + \frac{n}{2} \mod n$, and \distduplicateseight sets $\VarArray[i] = i^8+\frac{n}{2} \mod n$.
  We ran benchmarks with $64$-bit floating point elements and \emph{\pair}, \emph{\quartet}, and \emph{\bytes} data types.
  \pair (\quartet) consists of one (three) $64$-bit floating point elements as key and one $64$-bit floating point element of associated information.
  \bytes consists of $10$~bytes as key and $90$~bytes of associated information.
  \quartet and \bytes are compared lexicographically.
  For $n<2^{30}$, we perform each measurement $15$ times and for $n\geq 2^{30}$, we perform each measurement twice.
Unless stated otherwise, we report the average over all runs and use $64$-bit floating point elements.

We ran our experiments on machines with one AMD Ryzen $+1800$ 8-core processor (\emph{\pcamd}), two Intel Xeon E5-2683 v4 16-core processors (\emph{\pcinteltwo}), and four Intel Xeon E5-4640 8-core processors (\emph{\pcintelfour}).
\pcinteltwo and \pcintelfour are equipped with $512$~GiB of memory, \pcamd is equipped with $32$~GiB of memory.
We use the \texttt{taskset} tool to set the CPU affinity for speedup benchmarks.
We tested all parallel algorithms on  \distuniform input with and without hyper-threading.
Hyper-threading did not slow down any algorithm.
Thus, we give results of all algorithms with hyper-threading.
Overall, we executed more than $12\,000$ combinations of different algorithms, input distributions and sizes, data types and machines.
We now present a selection of our measurements and discuss our results.
For the remaining (detailed) running time and hardware counter measurements, we refer to \autoref{app:more measurements}.

\subparagraph*{Sequential Algorithms.}\label{sec:sequential comparison}

\begin{figure}[tbp]
  \begin{minipage}[t]{0.48\linewidth}%
    \hspace{-2mm}%
    \includegraphics[]{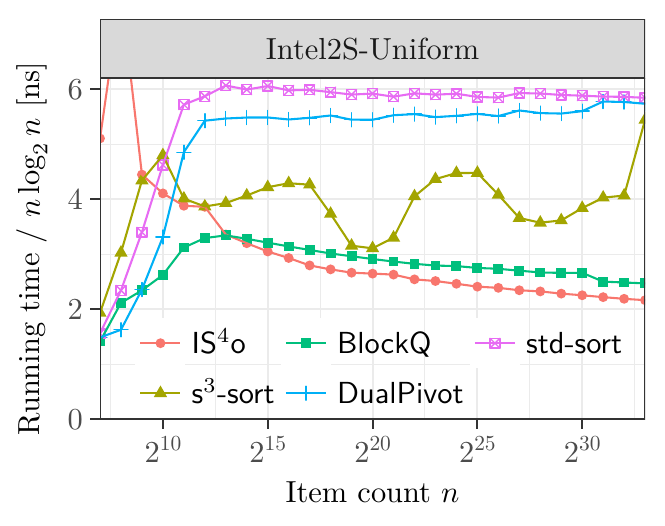}%
  \caption{\label{fig:m2 sequential random}%
    Running times of sequential algorithms on input distribution \distuniform executed on machine \pcinteltwo.
  }\end{minipage}\hfill%
  \begin{minipage}[t]{0.49\linewidth}%
    \hspace{-4mm}%
    \includegraphics[]{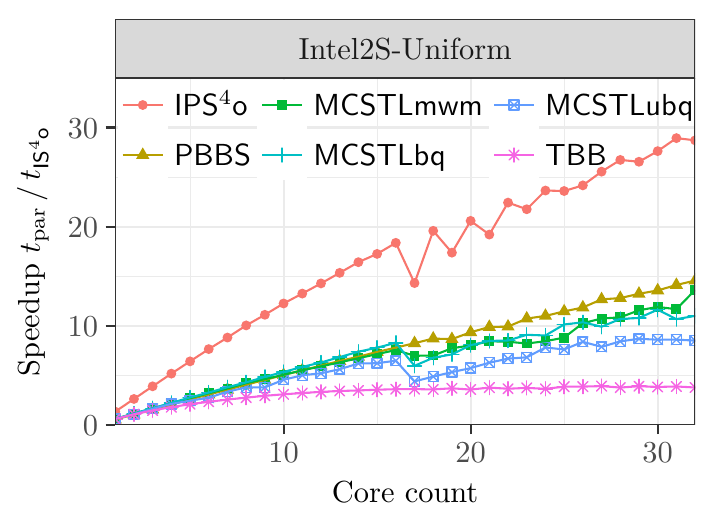}%
  \caption{\label{fig:m2_speedup}%
    Speedup of parallel algorithms with different number of cores relative to our sequential implementation \algoissssort on \pcinteltwo, sorting $2^{30}$ elements of input distribution \distuniform.
  }\end{minipage}%
\end{figure}

\autoref{fig:m2 sequential random} shows the running times of sequential algorithms on  \distuniform input executed on machine~\pcinteltwo.
We see that \algoissssort is faster than its closest competitor, \algosblock,
by a factor of $1.14$ for $n = 2^{32}$.
On machine~\pcintelfour (\pcamd), \algoissssort outperforms \algosblock even by a factor of $1.22$ ($1.57$).
\algosyaros and \algossort, which do not avoid branch mispredictions, are at least a factor of $1.86$ slower than \algoissssort for $n=2^{32}$.
The number of branch mispredictions of these algorithms for this input size is about $10$ times larger than that of \algoissssort.
\algossssort is the slowest sequential sorting algorithm avoiding branch mispredictions and has fluctuations in running time for varying input sizes.
Due to the initial overhead, \algoissssort is slower than \algosblock for $n \leq 2^{15}$.

As expected, the running times for inputs with a moderate number of different keys (\distduplicatestwice)
are similar to the running times for \distuniform.
When the number of different keys decreases (\distexpo, \distduplicateseight, and \distduplicatesroot in decreasing order),
  \algoissssort becomes even faster by a factor of up to two on all machines.
The running times of the competitors also decrease.
However, only \algosyaros on \pcinteltwo with \distduplicatesroot distributed input comes close for $n\geq 2^{28}$.
Only input \distones and (almost) sorted input are hard for \algoissssort; for example, \algosyaros outperforms \algoissssort on \distalmostsorted input by a factor of $1.70$ for $n=2^{32}$ (\pcinteltwo).
For detailed measurements see also \autoref{fig:sequential random collection}-\ref{fig:sequential input distribution 133} in \autoref{app:more measurements}.

\subparagraph*{Parallel Algorithms.}\label{sec:parallel comparison}
\begin{figure}[tbp]
  \centering%
  \includegraphics[]{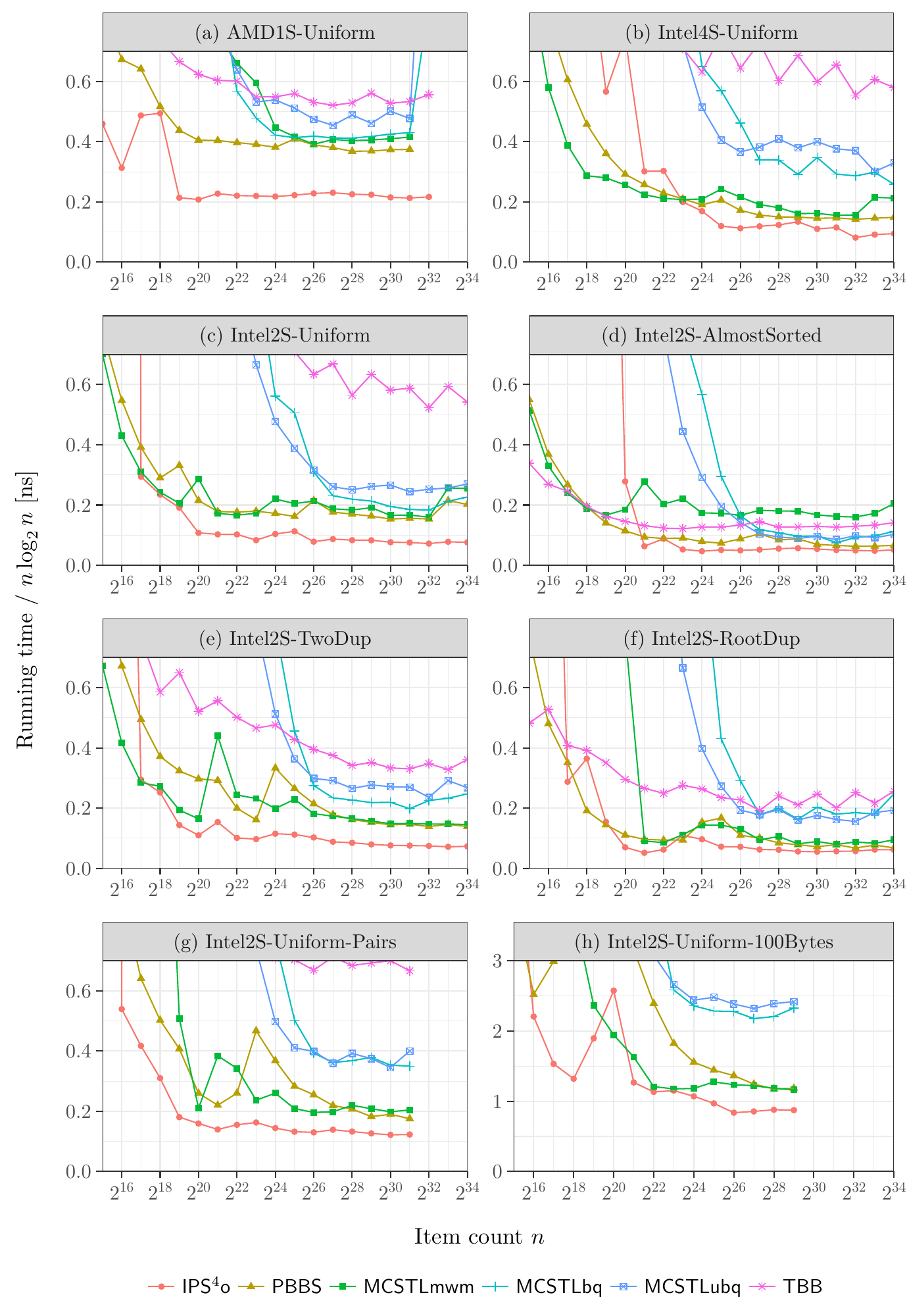}%
  \caption{\label{fig:32selection_mx_random}%
    Running times of parallel algorithms on different input distributions executed on different machines.
  }%
\end{figure}

\autoref{fig:32selection_mx_random}~(a-c) presents experiments of parallel algorithms on different machines for \distuniform input.
We see that \algoiparassssort outperforms its closest competitors, e.g., for $n=2^{32}$ on \pcinteltwo (\pcamd) by a factor of $2.13$ ($1.75$),
  and all but \algoptbb and \algoiparassssort fail to sort this input size on \pcamd due to memory limitations.
For $n\geq 2^{26}$, \algoiparassssort outperforms its closest non-in-place competitors on \pcinteltwo (\pcamd) on average by a factor of $2.26$ ($1.69$) and its closest in-place competitors by a factor of $2.78$ ($1.98$).
For the same input sizes, \algoiparassssort outperforms its closest competitors on \pcintelfour in average just by a factor of $1.41$.
We believe that the small difference in running time between \algoiparassssort and its competitors on \pcintelfour is caused by two factors:
  The slower memory modules (\textsf{DDR4} vs. \textsf{DDR3}), and the long load delays due to a ring interconnect between four sockets.

In \autoref{fig:32selection_mx_random}~(d-e), we present running times of parallel algorithms on input distributions with duplicates (\distduplicatestwice and \distduplicatesroot) on machine \pcinteltwo.
For $n\geq 2^{26}$ and a moderate number of different keys (\distduplicatestwice), \algoiparassssort still outperforms its in-place competitors on average
  by a factor of at least $2.88$ and its non-in-place competitors on average by a factor of at least $1.91$.
Experiments have shown that the running times on \distduplicateseight and \distexpo are similar to the running times on \distduplicatestwice.
We also see that the non-in-place algorithms become almost as fast as \algoiparassssort if we sort inputs which contain few different keys (\distduplicatesroot).
However, \algoiparassssort still outperforms its in-place competitors by a factor of at least $3.43$ on this input for $n\geq2^{20}$.
\autoref{fig:32selection_mx_random}~(f) depicts the running times of parallel algorithms on \distalmostsorted distributions on \pcinteltwo.
On \distalmostsorted and \distreversesorted, the fastest non-in-place algorithm, \algoppbbs, performs similarly to \algoiparassssort for large input sizes.
Only on \distsorted and \distones, \algoiparassssort is outperformed by \algoptbb, an in-place competitor.
This is because \algoptbb detects these pre-sorted input distributions and terminates immediately.
Further benchmarks on machines \pcintelfour and \pcamd show that \algoiparassssort also outperforms its non-in-place competitors on any machine and
  that \algoiparassssort is much faster than its in-place competitors except in the case of \distsorted and \distones inputs.
For detailed measurements see also \autoref{fig:parallel input distribution 132}-\ref{fig:parallel input distribution 133} in \autoref{app:more measurements}.

In \autoref{fig:32selection_mx_random}~(g-h), we give running times of \pair and \bytes data types on machine \pcinteltwo with uniformly distributed keys.
We see that \algoiparassssort outperforms its competitors, e.g., by a factor of $1.33$ (non-in-place competitor) and by a factor of $2.67$ (its in-place competitor) for $2^{29}$  \bytes elements.
Further benchmarks on machines \pcintelfour and \pcamd show similar running times.
For detailed measurements see also \autoref{fig:parallel data type 132}-\ref{fig:parallel data type 133} in \autoref{app:more measurements}.

\autoref{fig:m2_speedup} depicts the speedup of parallel algorithms executed on different numbers of cores relative to our sequential implementation \algoissssort on \pcinteltwo,
  sorting \distuniform input ($n=2^{30}$).
We see that \algoiparassssort outperforms its competitors on any number of cores.
\algoiparassssort outperforms \algoissssort on 32 cores by a factor of $28.71$, whereas its fastest non-in-place competitor, \algoppbbs, outperforms \algoissssort just by a factor of $14.54$.
The in-place algorithms, \algopunbalancedsort and \algopbalancedsort, scale similarly to \algoppbbs up to $16$ cores but begin lagging behind for larger numbers of cores.
Further measurements show that \algoiparassssort scales similarly on \pcamd.
On \pcintelfour, \algoiparassssort scales well on the first processor. However, as the input data is stored in the memory of the first processor, adding the second, third and fourth processors speeds up \algoiparassssort by an additional factor of only 1.45; again caused by the slower memory modules (\textsf{DDR4} vs. \textsf{DDR3}) and the long load delays due to a ring interconnect between four sockets.
For detailed measurements see also \autoref{fig:speedup collection} in \autoref{app:more measurements}.

\section{Conclusion and Future Work}\label{s:conclusion}

In-place super scalar samplesort (\algoiparassssort) is among the fastest comparison-based
sorting algorithms both sequentially and on multi-core machines.
The algorithm can also be used for data distribution and local sorting in distributed memory parallel algorithms (e.g.,~\cite{ABSS15}).
Somewhat surprisingly, there is even an advantage over non-in-place algorithms because \algoiparassssort  saves on overhead for memory allocation, associativity misses and write allocate misses.
Compared to previous parallel in-place algorithms, improvements by more than a factor of two are possible.
The main case where \algoiparassssort is slower than the best competitors (\algossssort and BlockQuicksort)
  is for sequentially sorting large objects (\quartet and \bytes, see \autoref{app:more measurements})
  because \algoiparassssort moves elements twice in one distribution step.
In this case, the overhead for the oracle information of \algossssort is small and we could try an almost-in-place variant of \algossssort with element-wise in-place permutation.

Several improvements of \algoiparassssort\ can be considered.
Besides careful adaptation of parameters like $\VarBucketCount$, $\BlockSize$, $\OversamplingFactor$, and the choice of base case algorithm,
  one would like to avoid contention on the bucket pointers in the block permutation phase when $\VarThreadCount$ is large.
Perhaps the most important improvement would be to make \algoiparassssort\ aware of non-uniform memory access costs (NUMA) depending on the memory module holding a particular piece of data. This can be done by preferably assigning pieces of the input array to ``close-by'' cores both for local classification and when switching to sequential sorting.
In situations with little NUMA effects, we could
ensure that our data blocks correspond to pages of the virtual memory.
Then, one can replace block permutation with relabelling the virtual memory addresses of the corresponding pages.

Coming back to the original motivation for an alternative to quicksort variants in standard libraries, we see \algoiparassssort\ as an interesting candidate. The main remaining issue is the code complexity.
When code size matters (e.g., as indicated by a compiler flag like {\tt -Os}), quicksort should still be used. Formal verification of the correctness of the implementation might help to increase trust in the remaining cases.

\subparagraph*{Acknowledgements.}
We would like to thank the authors of~\cite{shun2012brief, edelkamp2016blockquicksort} for sharing their code for evaluation.
Timo Bingmann and Lorenz Hübschle-Schneider~\cite{Lorenz2016ssss} kindly provided code that was used as a starting point for our implementation.

\bibliography{paper}

\begin{thebibliography}{10}

\bibitem{AGNS08}
Lars Arge, Michael~T Goodrich, Michael Nelson, and Nodari Sitchinava.
\newblock Fundamental parallel algorithms for private-cache chip
  multiprocessors.
\newblock In {\em 20th Symposium on Parallelism in Algorithms and Architectures
  (SPAA)}, pages 197--206. ACM, 2008.

\bibitem{ABSS15}
Michael Axtmann, Timo Bingmann, Peter Sanders, and Christian Schulz.
\newblock Practical massively parallel sorting.
\newblock In {\em 27th ACM Symposium on Parallelism in Algorithms and
  Architectures, (SPAA)}, 2015.
\newblock \href {http://dx.doi.org/10.1145/2755573.2755595}
  {\path{doi:10.1145/2755573.2755595}}.

\bibitem{bingmann2013parallel}
Timo Bingmann and Peter Sanders.
\newblock Parallel string sample sort.
\newblock In {\em European Symposium on Algorithms}, pages 169--180. Springer,
  2013.

\bibitem{blelloch2010low}
Guy~E Blelloch, Phillip~B Gibbons, and Harsha~Vardhan Simhadri.
\newblock Low depth cache-oblivious algorithms.
\newblock In {\em Proceedings of the twenty-second annual ACM symposium on
  Parallelism in algorithms and architectures}, pages 189--199. ACM, 2010.

\bibitem{blelloch1991comparison}
Guy~E Blelloch, Charles~E Leiserson, Bruce~M Maggs, C~Greg Plaxton, Stephen~J
  Smith, and Marco Zagha.
\newblock A comparison of sorting algorithms for the connection machine {CM-2}.
\newblock In {\em Proceedings of the third annual ACM symposium on Parallel
  algorithms and architectures}, pages 3--16. ACM, 1991.

\bibitem{BFV04}
G.~S. Brodal, R.~Fagerberg, and K.~Vinther.
\newblock Engineering a cache-oblivious sorting algorithm.
\newblock In {\em 6th Workshop on Algorithm Engineering and Experiments}, 2004.

\bibitem{CHEN200634}
Jing-Chao Chen.
\newblock A simple algorithm for in-place merging.
\newblock {\em Information Processing Letters}, 98(1):34 -- 40, 2006.
\newblock URL:
  \url{http://www.sciencedirect.com/science/article/pii/S002001900500339X},
  \href {http://dx.doi.org/http://dx.doi.org/10.1016/j.ipl.2005.11.018}
  {\path{doi:http://dx.doi.org/10.1016/j.ipl.2005.11.018}}.

\bibitem{cho2015paradis}
Minsik Cho, Daniel Brand, Rajesh Bordawekar, Ulrich Finkler, Vincent
  Kulandaisamy, and Ruchir Puri.
\newblock {PARADIS}: an efficient parallel algorithm for in-place radix sort.
\newblock {\em Proceedings of the VLDB Endowment}, 8(12):1518--1529, 2015.

\bibitem{edelkamp2016blockquicksort}
Stefan Edelkamp and Armin Weiss.
\newblock {BlockQuicksort}: Avoiding branch mispredictions in quicksort.
\newblock In {\em 24th European Symposium on Algorithms (ESA)}, volume~57 of
  {\em LIPIcs}, 2016.

\bibitem{Fran04}
Gianni Franceschini.
\newblock Proximity mergesort: Optimal in-place sorting in the cache-oblivious
  model.
\newblock In {\em Proceedings of the Fifteenth Annual ACM-SIAM Symposium on
  Discrete Algorithms}, SODA '04, pages 291--299, Philadelphia, PA, USA, 2004.
  Society for Industrial and Applied Mathematics.
\newblock URL: \url{http://dl.acm.org/citation.cfm?id=982792.982833}.

\bibitem{FraGef05}
Gianni Franceschini and Viliam Geffert.
\newblock An in-place sorting with {O(N log N)} comparisons and {O(N)} moves.
\newblock {\em J. ACM}, 52(4):515--537, July 2005.
\newblock URL: \url{http://doi.acm.org/10.1145/1082036.1082037}, \href
  {http://dx.doi.org/10.1145/1082036.1082037}
  {\path{doi:10.1145/1082036.1082037}}.

\bibitem{FraMck70}
W.~D. Frazer and A.~C. McKellar.
\newblock Samplesort: A sampling approach to minimal storage tree sorting.
\newblock {\em J. ACM}, 17(3):496--507, July 1970.
\newblock URL: \url{http://doi.acm.org/10.1145/321592.321600}, \href
  {http://dx.doi.org/10.1145/321592.321600} {\path{doi:10.1145/321592.321600}}.

\bibitem{Geffert2009}
Viliam Geffert and Jozef Gajdo{\v{s}}.
\newblock Multiway in-place merging.
\newblock In Miros{\l}aw Kuty{\l}owski, Witold Charatonik, and Maciej
  G{\k{e}}bala, editors, {\em 17th Symposium on Fundamentals of Computation
  Theory (FCT)}, volume 5699 of {\em LNCS}, pages 133--144. Springer, 2009.
\newblock URL: \url{http://dx.doi.org/10.1007/978-3-642-03409-1_13}, \href
  {http://dx.doi.org/10.1007/978-3-642-03409-1_13}
  {\path{doi:10.1007/978-3-642-03409-1_13}}.

\bibitem{hoare1962quicksort}
Charles~AR Hoare.
\newblock Quicksort.
\newblock {\em The Computer Journal}, 5(1):10--16, 1962.

\bibitem{Lorenz2016ssss}
Lorenz H{\"u}bschle-Schneider.
\newblock Super scalar sample sort.
\newblock \url{https://github.com/lorenzhs/ssssort}, retrieved September 15,
  2016.

\bibitem{jurkiewicz2015model}
Tomasz Jurkiewicz and Kurt Mehlhorn.
\newblock On a model of virtual address translation.
\newblock {\em Journal of Experimental Algorithmics (JEA)}, 19, 2015.

\bibitem{KS06}
K.~Kaligosi and P.~Sanders.
\newblock How branch mispredictions affect quicksort.
\newblock In {\em 14th European Symposium on Algorithms (ESA)}, volume 4168 of
  {\em LNCS}, pages 780--791, 2006.

\bibitem{kaligosi2006branch}
Kanela Kaligosi and Peter Sanders.
\newblock How branch mispredictions affect quicksort.
\newblock In {\em European Symposium on Algorithms}, pages 780--791. Springer,
  2006.

\bibitem{katajainen1996practical}
Jyrki Katajainen, Tomi Pasanen, and Jukka Teuhola.
\newblock Practical in-place mergesort.
\newblock {\em Nord. J. Comput.}, 3(1):27--40, 1996.

\bibitem{kokot2017even}
Marek Kokot, Sebastian Deorowicz, and Maciej Dlugosz.
\newblock Even faster sorting of (not only) integers.
\newblock {\em arXiv preprint arXiv:1703.00687}, 2017.

\bibitem{KLMQ14}
Shrinu Kushagra, Alejandro L\'{o}pez-Ortiz, J.~Ian Munro, and Aurick Qiao.
\newblock Multi-pivot quicksort: Theory and experiments.
\newblock In {\em Meeting on Algorithm Engineering \& Experiments (ALENEX)},
  pages 47--60, Philadelphia, PA, USA, 2014. AMS.
\newblock URL: \url{http://dl.acm.org/citation.cfm?id=2790174.2790180}.

\bibitem{MehSan03}
K.~Mehlhorn and P.~Sanders.
\newblock Scanning multiple sequences via cache memory.
\newblock {\em Algorithmica}, 35(1):75--93, 2003.

\bibitem{musser1997introspective}
David~R. Musser.
\newblock Introspective sorting and selection algorithms.
\newblock {\em Softw., Pract. Exper.}, 27(8):983--993, 1997.

\bibitem{Rah03}
N.~Rahman.
\newblock {\em Algorithms for Memory Hierarchies}, volume 2625 of {\em LNCS},
  chapter Algorithms for Hardware Caches and TLB, pages 171--192.
\newblock Springer, 2003.

\bibitem{reinders2007intel}
James Reinders.
\newblock {\em Intel threading building blocks: outfitting C++ for multi-core
  processor parallelism}.
\newblock " O'Reilly Media, Inc.", 2007.

\bibitem{SanWas11}
Peter Sanders and Jan Wassenberg.
\newblock Engineering a multi-core radix sort.
\newblock In {\em 17th Euro-Par Conference}, volume 6853 of {\em LNCS}, pages
  160--169. Springer, 2011.

\bibitem{sanders2004super}
Peter Sanders and Sebastian Winkel.
\newblock Super scalar sample sort.
\newblock In {\em 12th European Symposium on Algorithms (ESA)}, volume 3221 of
  {\em LNCS}, pages 784--796. Springer, 2004.

\bibitem{shun2012brief}
Julian Shun, Guy~E Blelloch, Jeremy~T Fineman, Phillip~B Gibbons, Aapo Kyrola,
  Harsha~Vardhan Simhadri, and Kanat Tangwongsan.
\newblock Brief announcement: the problem based benchmark suite.
\newblock In {\em Proceedings of the twenty-fourth annual ACM symposium on
  Parallelism in algorithms and architectures}, pages 68--70. ACM, 2012.

\bibitem{putze2007mcstl}
J.~Singler, P.~Sanders, and F.~Putze.
\newblock {MCSTL}: The multi-core standard template library.
\newblock In {\em 13th Euro-Par Conference}, volume 4641 of {\em LNCS}, pages
  682--694. Springer, 2007.
\newblock URL: \url{http://dx.doi.org/10.1007/978-3-540-74466-5_72}.

\bibitem{TsiZha03}
P.~Tsigas and Y.~Zhang.
\newblock A simple, fast parallel implementation of quicksort and its
  performance evaluation on {SUN} {Enterprise} 10000.
\newblock In {\em PDP}, pages 372--381. IEEE Computer Society, 2003.
\newblock URL:
  \url{http://doi.ieeecomputersociety.org/10.1109/EMPDP.2003.1183613}.

\bibitem{yaroslavskiy2009dual}
Vladimir Yaroslavskiy.
\newblock Dual-pivot quicksort.
\newblock {\em Research Disclosure}, 2009.

\end{thebibliography}

\newpage
\appendix

\section{Details of the Algorithm}\label{app:algorithm}
\subparagraph*{Empty block movement}
An important observation is that within each stripe, all full blocks are at the beginning, followed by all empty blocks.
This arrangement fulfils the invariant used during permutation, which is why there is no need to move empty blocks in the sequential algorithm.
It also means that in the parallel algorithm, only the buckets crossing a stripe boundary need to be fixed.

To do so, each thread finds the bucket that starts before the end of its stripe, but ends after it.
It then finds the stripe in which that bucket ends (which will be the following stripe in most cases) and
  moves the last full block in the bucket into the first empty block in the bucket.
It continues to do this until either all empty blocks in its stripe are filled or all full blocks in the bucket have been moved.

In rare cases, very large buckets exist that cross multiple stripes.
In this case, each thread will first count how many blocks in the preceding stripes need to be filled.
It will then skip that many blocks at the end of the bucket before starting to fill its own empty blocks.

\section{Details of the Analysis}\label{app:analysis}

\begin{proof}[Proof of \autoref{thm:io}]
  It can be shown using Chernoff bounds that an oversampling ratio of
  $\alpha=\Om{\log kt}$ is sufficient to produce (non-equality) buckets
  of size $\Oh{\frac{N}{k}}$ with high probability for subproblems of size $N$. Hence, $\Oh{\log_k \frac{n}{n_0}}$
  levels of recursion suffice with high probability. On the other hand,
  for $\alpha=\Oh{t}$, even sequentially processing the sample does not
  become a bottleneck.

  During the block distribution phase, each thread reads $\Oh{\frac{n}{tb}}$ logical data blocks, writes them
  to the buffers in its private cache, and eventually moves them back to main memory.

  The same asymptotic cost occurs for moving blocks during block permutation.
  Each thread performs $\Oh{\frac{n}{tb}}$ successful acquisitions of the next block in a bucket.
  Charging $\Oh{t}$ I/Os for this accounts for possible contention with other
  threads. Overall, we get cost $\Oh{\frac{n}{b}}=\Oh{\frac{n}{tB}}$.
  Similarly, there are $k$ unsuccessful acquisitions before termination is
  determined, for which we charge an overall cost of $\Oh{kt}$ I/Os.
  Since we assume $n=\Om{kt^2B}$, we have $k=\Oh{\frac{n}{t^2B}}$ and hence $\Oh{kt}=\Oh{\frac{n}{tB}}$.

  In the cleanup phase, we consider a case distinction with respect to $k$ and $t$.
  If $k\leq t$, then each thread processes at most one bucket and it has to move
  elements from $t+2$ distribution buffers and bucket boundaries.
  This amounts to a cost of $\Oh{\frac{tb}{B}}=\Oh{t^2}$ I/Os. Since $n=\Om{t^3B}$,
  we get $t^2=\Oh{\frac{n}{tB}}$.
  If $k>t$, then each thread processes $\Oh{\frac{k}{t}}$ buckets with a total cost
  of $\Oh{k/t\cdot t^2}=\Oh{kt}$. Since $n=\Om{kt^2B}$, we have $kt=\Oh{\frac{n}{tB}}$.
\end{proof}

\subparagraph*{Comparing the I/O volume of \algoissssort\ and \algossssort.}
Both algorithms read and write the data once for the base case -- $16n$ bytes (of I/O volume).
\algoissssort\ reads and writes all data both during data distribution and block permutation phase -- $32n$ bytes or $48n$ bytes overall.
\algossssort\ reads the element twice but writes them only once in its distribution algorithm -- $24n$ bytes.
This sounds like a slight advantage. However, now we come to overheads unique to \algossssort.
First, the algorithm reads and writes an oracle sequence that indicates the bucket for each element -- $2n$ bytes.
\algossssort\ has to copy the sorted result data back to the input array -- $16n$ bytes.
It also has to allocate the temporary arrays. For security reasons, that memory is zeroed by the operating system -- $9n$ bytes.%
\footnote{In current versions of the Linux kernel this is done by a single thread and thus results in a huge scalability bottleneck.}
When writing to the temporary arrays or during copying back, there are so called allocate misses that happen when an element is written to a cache block that is currently not in memory -- that block is read to the cache because the CPU does not know that none of the data in that block will ever be read.
This amounts to an I/O volume of up to $17n$ bytes. Furthermore, \algossssort\ may suffer more associativity misses than \algoissssort\ -- the relative positions of the buckets in the temporary array are not coordinated while
\algoissssort\ essentially sweeps a window of size $\approx bk$ through the memory during the distribution phase.
For an average case analysis refer to \cite{MehSan03}. Even ignoring the latter overhead we get a total I/O volume of $86n$ byte -- more than twice as much as \algoissssort.
Much of this overhead can be reduced using measures that are non-portable (or hard to make portable).
In particular, non-temporal writes eliminate the allocation misses and also help to eliminate the associativity misses.
One could also use a base case sorter that does the copying back as as side-effect when the number of recursion levels is odd.
When sorting multiple times within an application, one can keep the temporary arrays without having to reallocate them. However, this may require a different interface to the sorter.
Overall, depending on many implementation details, \algoissssort\ may require slightly more I/O volume than \algossssort\ or significantly less.

\begin{proof}[Proof of \autoref{thm:space}]
  The main space overhead is for $k$ buffer blocks of size $b$ for each of $t$ threads.
  This bound also covers smaller amounts of memory for
  the search tree ($\Oh{k}$), swap buffers and overflow buffers ($\Oh{bt}$),
  read and write pointers ($\Oh{kB}$ if we avoid false sharing), end pointers, and bucket boundary pointers.
  All of these data structures can be used for all levels of recursion.
  The term $\Oh{\log_k \frac{n}{n_0}}$ stems from the space for the recursion stack itself.
\end{proof}

\section{More Measurements}\label{app:more measurements}

\begin{table}[h!]
  \centering
\begin{tabular}{lrr|rrrrr}
 Machine     & Algo              & Competitor   & \distuniform & \distexpo & Almost & \distduplicatesroot & \distduplicatestwice \\ \toprule
\pcinteltwo  & \algoissssort     & both         & 1.14         & 1.23      & 0.59              & 0.97                & 1.17                 \\
\pcintelfour & \algoissssort     & both         & 1.21         & 1.54      & 0.77              & 1.65                & 1.44                 \\
\pcamd       & \algoissssort     & both         & 1.57         & 2.02      & 0.65              & 1.37                & 1.17                 \\ \midrule
\pcinteltwo  & \algoiparassssort & in-place     & 2.54         & 3.43      & 1.88              & 2.73                & 3.02                 \\
             &                   & non-in-place & 2.13         & 1.79      & 1.29              & 1.19                & 1.86                 \\ \midrule
\pcintelfour & \algoiparassssort & in-place     & 3.52         & 4.35      & 3.62              & 3.19                & 2.89                 \\
             &                   & non-in-place & 1.75         & 1.69      & 1.84              & 1.15                & 1.19                 \\  \midrule
\pcamd       & \algoiparassssort & in-place     & 1.57         & 3.18      & 1.81              & 2.37                & 2.02                 \\
             &                   & non-in-place & \textsf{OOM}          & \textsf{OOM}       & \textsf{OOM}               & \textsf{OOM}                 & \textsf{OOM}                  \\
\end{tabular}
\caption{
  The first three rows show the speedups of \algoissssort relative to the fastest sequential in-place and non-in-place competitor on different input types executed on machine \pcinteltwo, \pcintelfour, and \pcamd for $n=2^{32}$.
  The last rows show the speedups of \algoiparassssort relative to the fastest parallel in-place and non-in-place competitor on different input types executed on different machine instances for $n=2^{32}$.
  Measurements in cells labeled with \textsf{OOM} ran out of memory.}
  \label{tab:seq parallel speedups}
\end{table}

\begin{figure}[tbp]
  \centering%
  \includegraphics[]{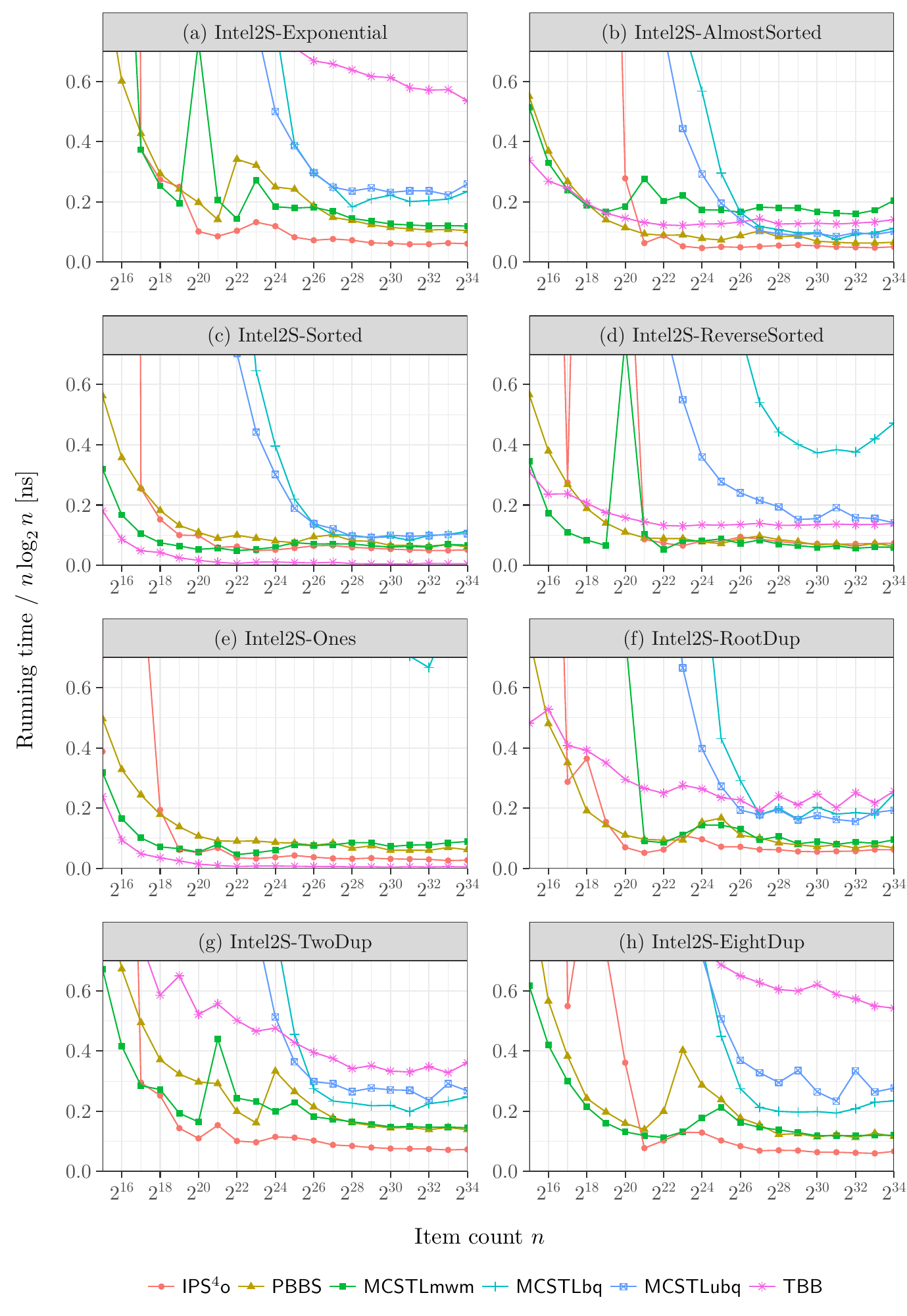}%
  \caption{\label{fig:parallel input distribution 132}%
    Running times of parallel algorithms on different input distributions executed on machine \pcinteltwo.}
\end{figure}
\begin{figure}[tbp]
  \centering%
  \includegraphics[]{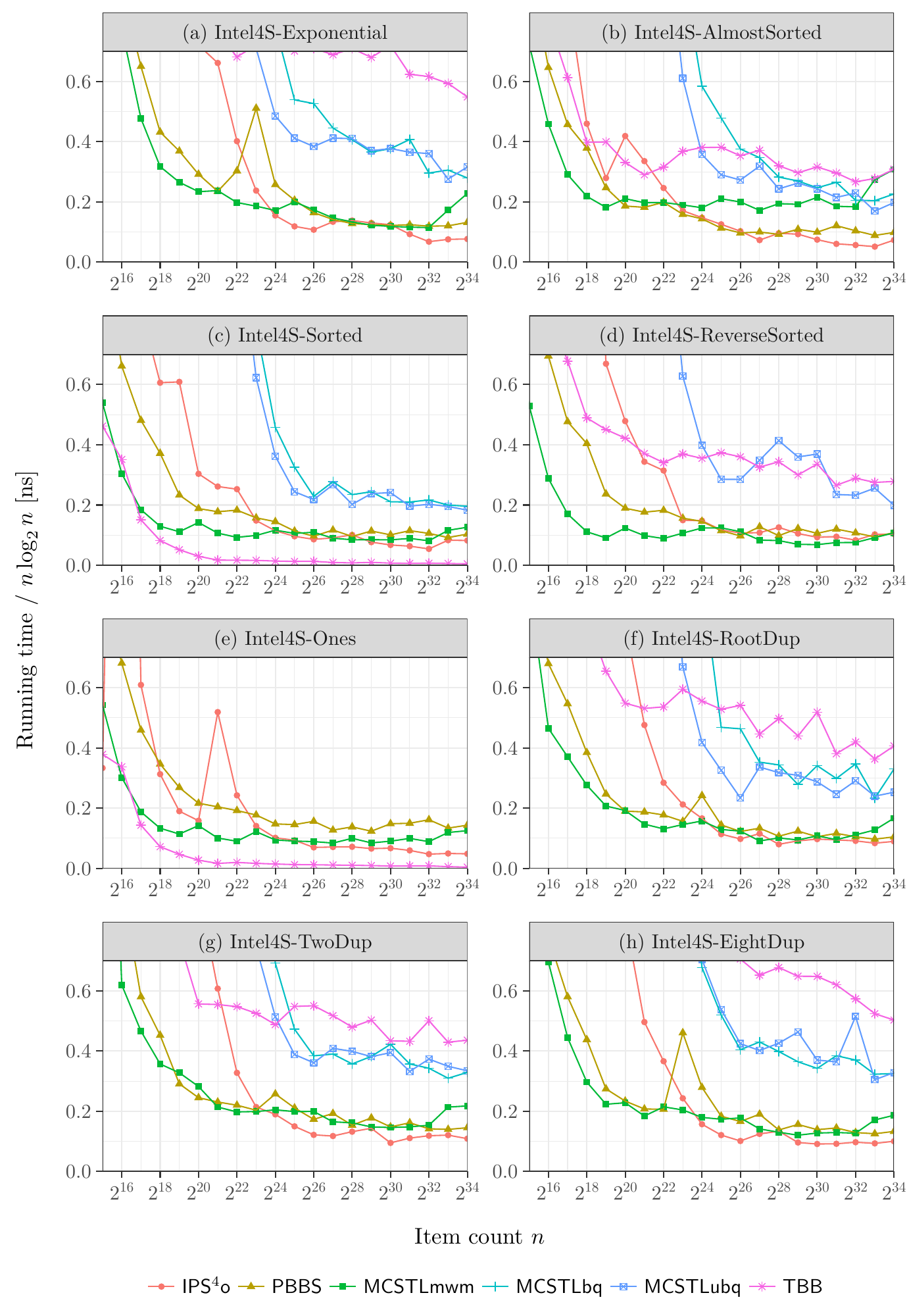}%
  \caption{\label{fig:parallel input distribution 127}%
    Running times of parallel algorithms on different input distributions executed on machine \pcintelfour.}
\end{figure}
\begin{figure}[tbp]
  \centering%
  \includegraphics[]{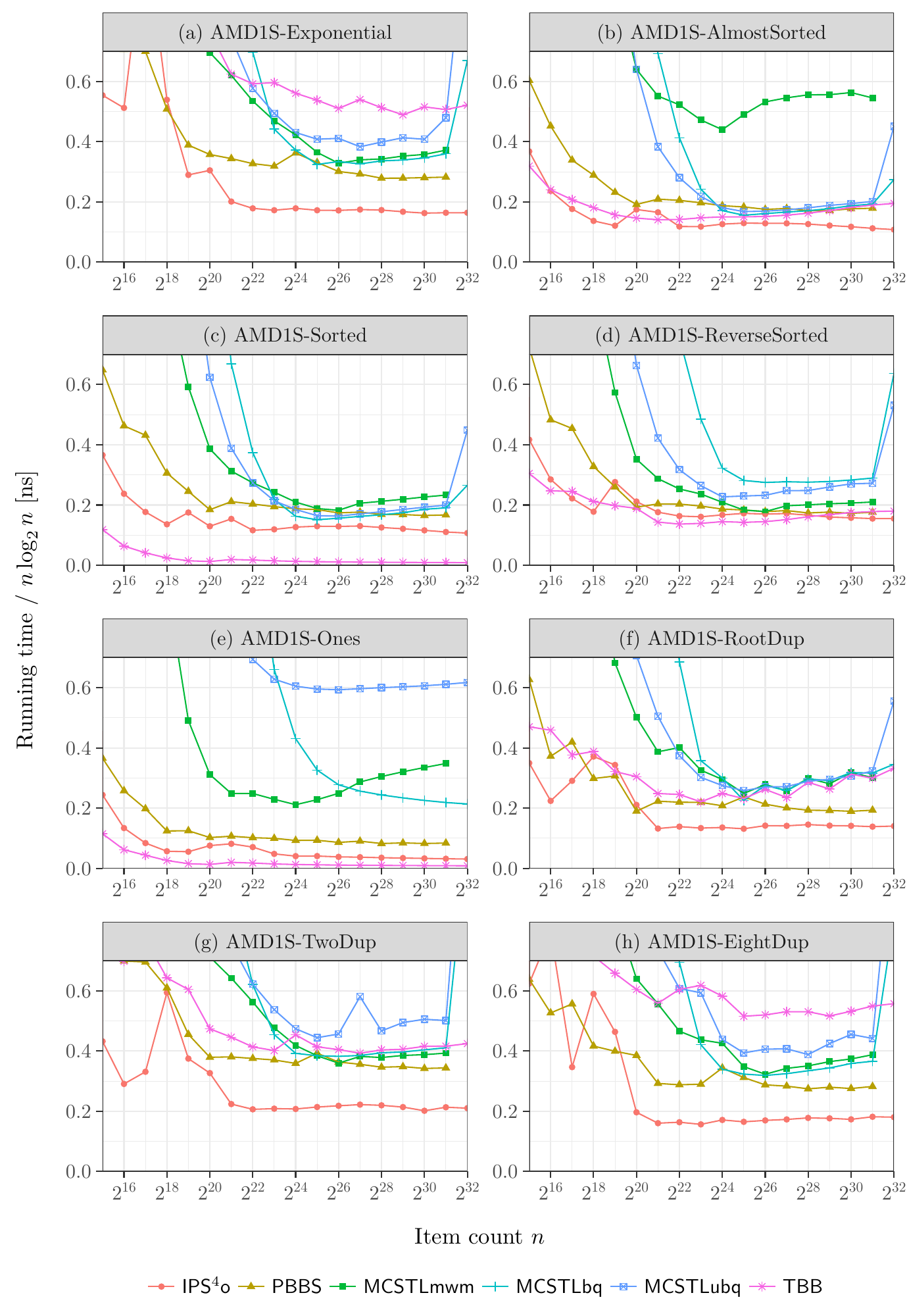}%
  \caption{\label{fig:parallel input distribution 133}%
    Running times of parallel algorithms on different input distributions executed on machine \pcamd.}
\end{figure}
\begin{figure}[tbp]
  \centering%
  \includegraphics[]{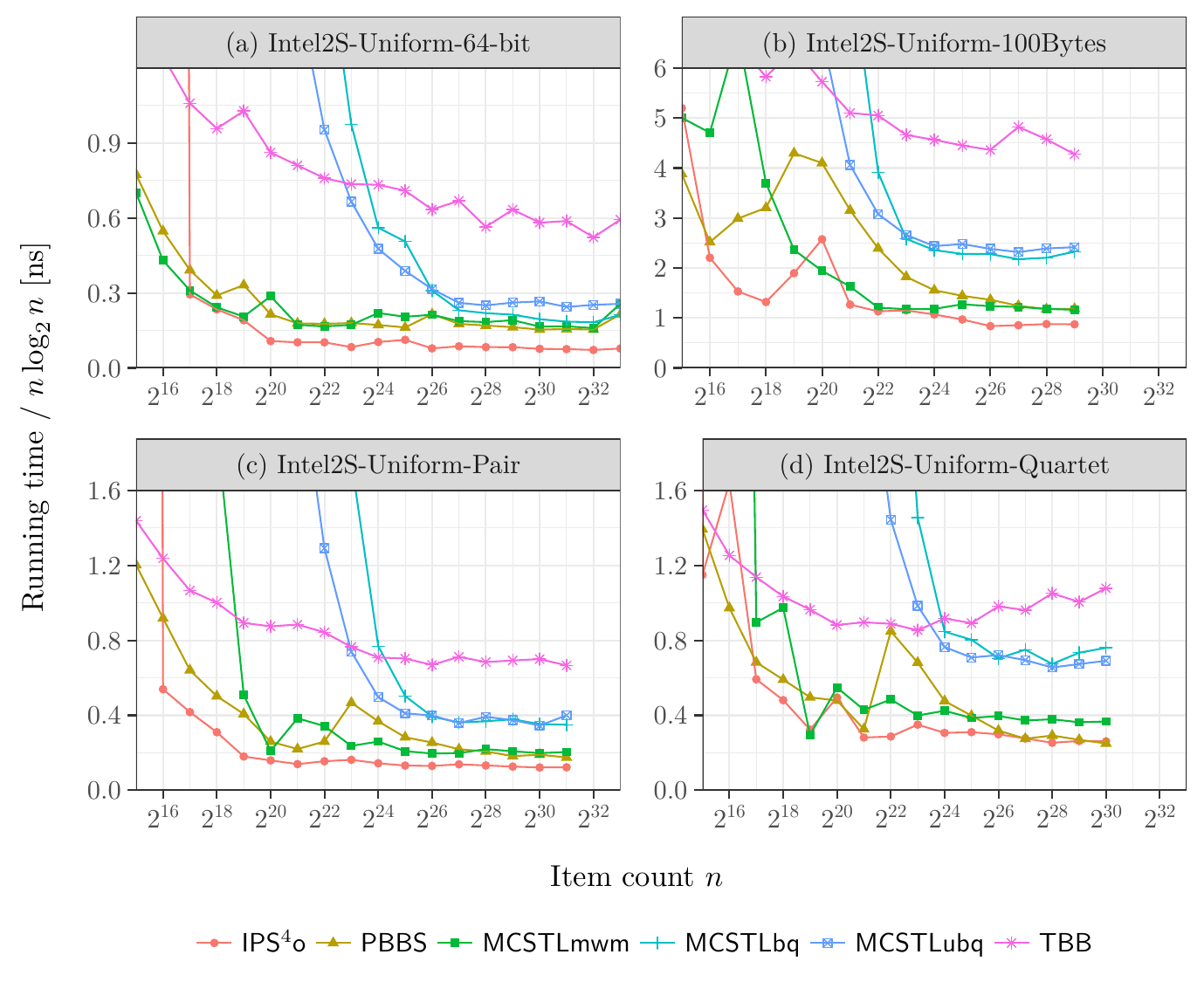}%
  \caption{\label{fig:parallel data type 132}%
    Running times of parallel algorithms on different data types of input distribution \distuniform executed on machine \pcinteltwo.}
\end{figure}
\begin{figure}[tbp]
  \centering%
  \includegraphics[]{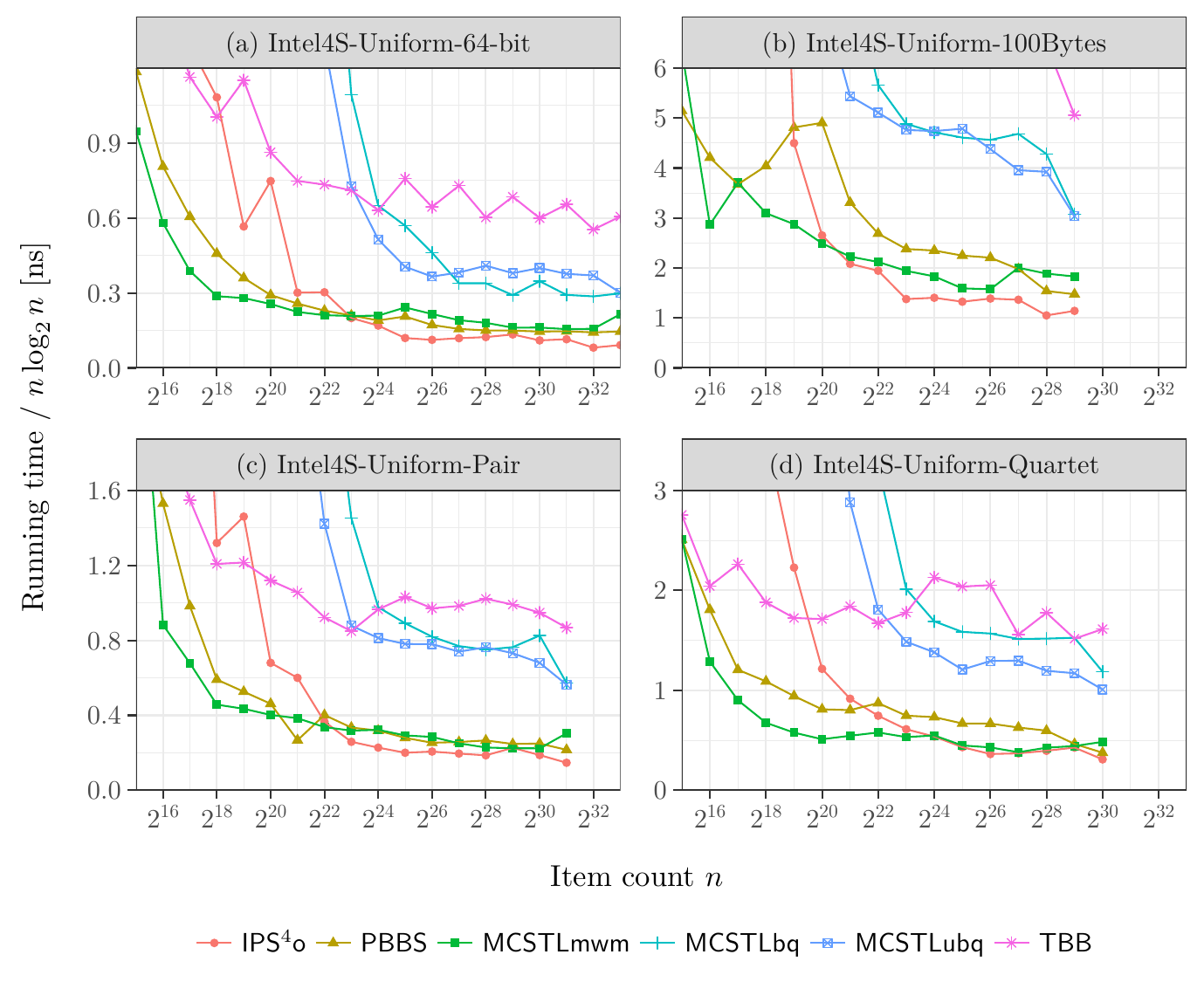}%
  \caption{\label{fig:parallel data type 127}%
    Running times of parallel algorithms on different data types of input distribution \distuniform executed on machine \pcintelfour.}
\end{figure}
\begin{figure}[tbp]
  \centering%
  \includegraphics[]{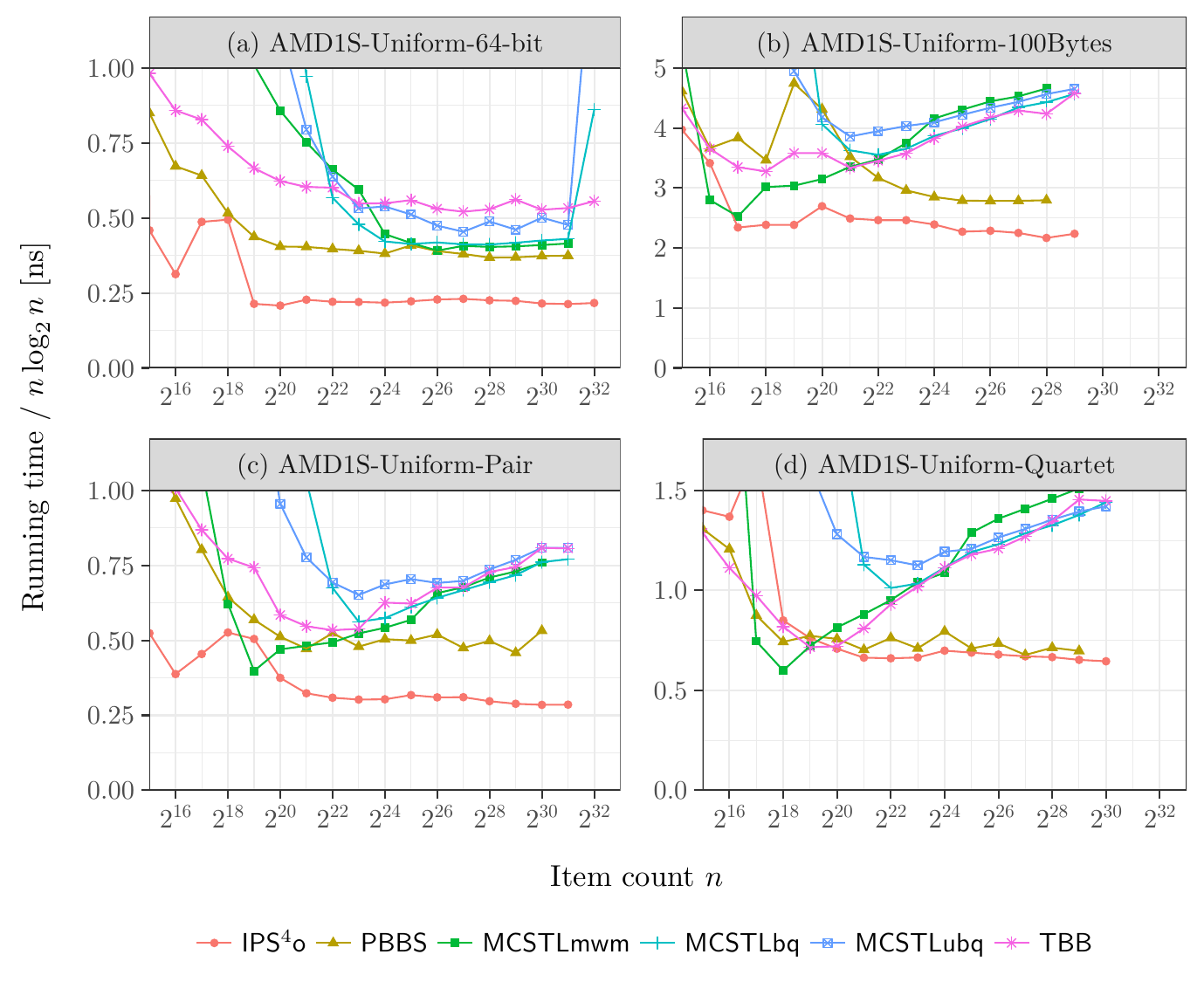}%
  \caption{\label{fig:parallel data type 133}%
    Running times of parallel algorithms on different data types of input distribution \distuniform executed on machine \pcamd.}
\end{figure}
\begin{figure}[tbp]
  \centering%
  \includegraphics[]{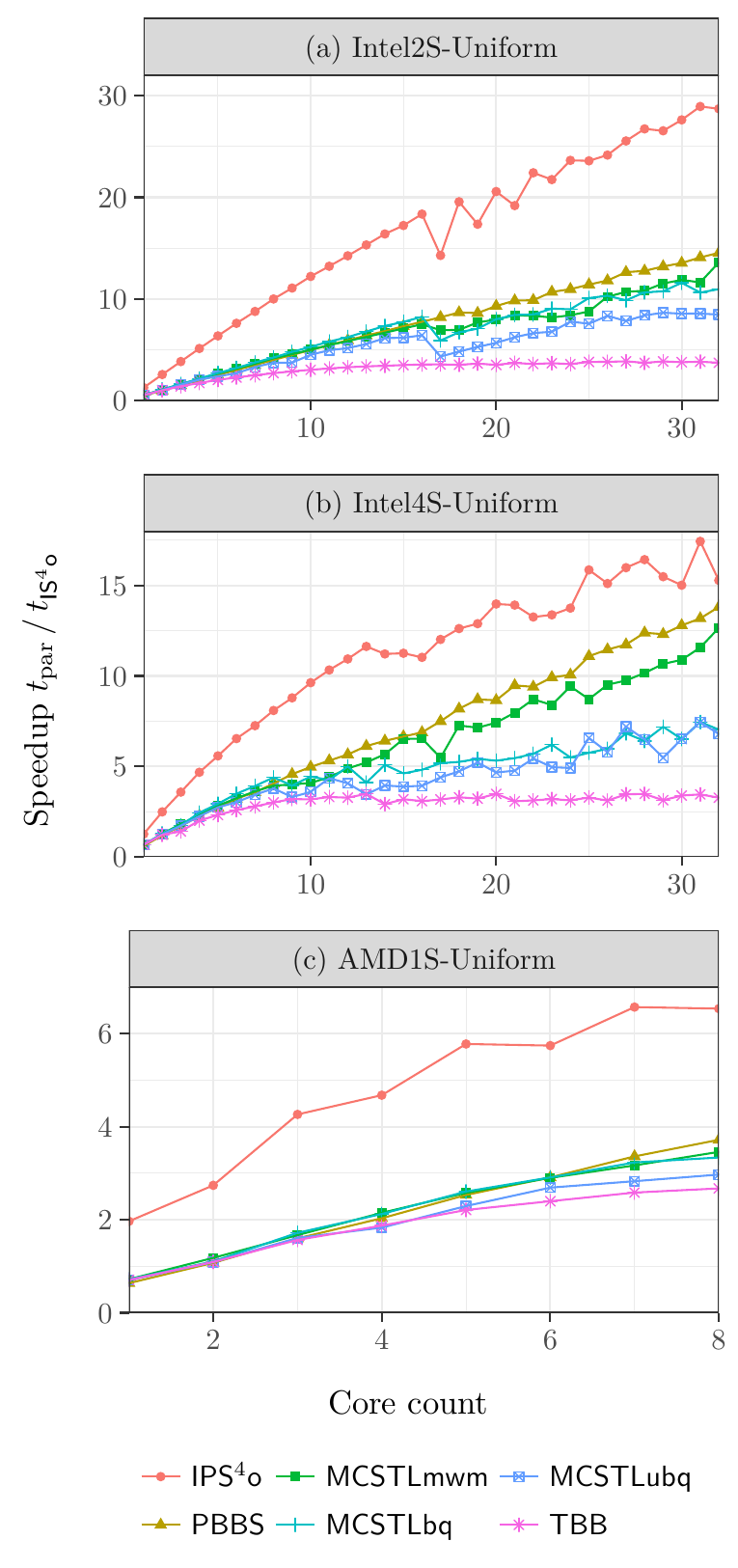}%
  \caption{\label{fig:speedup collection}%
    Speedup of parallel algorithms with different number of cores relative to our sequential implementation \algoissssort on machines, sorting $2^{30}$ elements of input distribution \distuniform.}
\end{figure}

\begin{figure}[tbp]
  \centering%
  \includegraphics[]{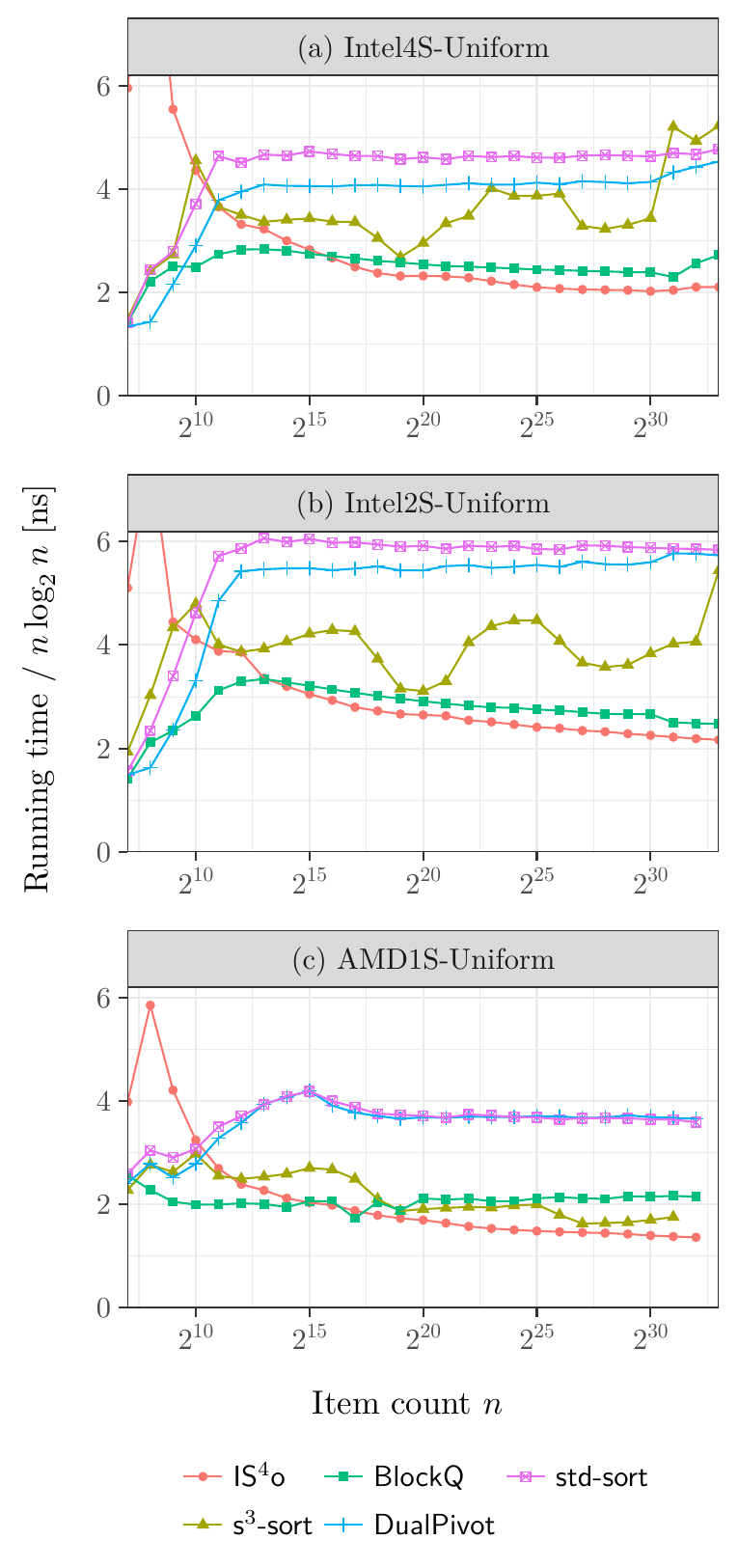}%
  \caption{\label{fig:sequential random collection}%
    Running times of sequential algorithms on input distribution \distuniform executed on different machines.}
\end{figure}
\begin{figure}[tbp]
  \centering%
  \includegraphics[]{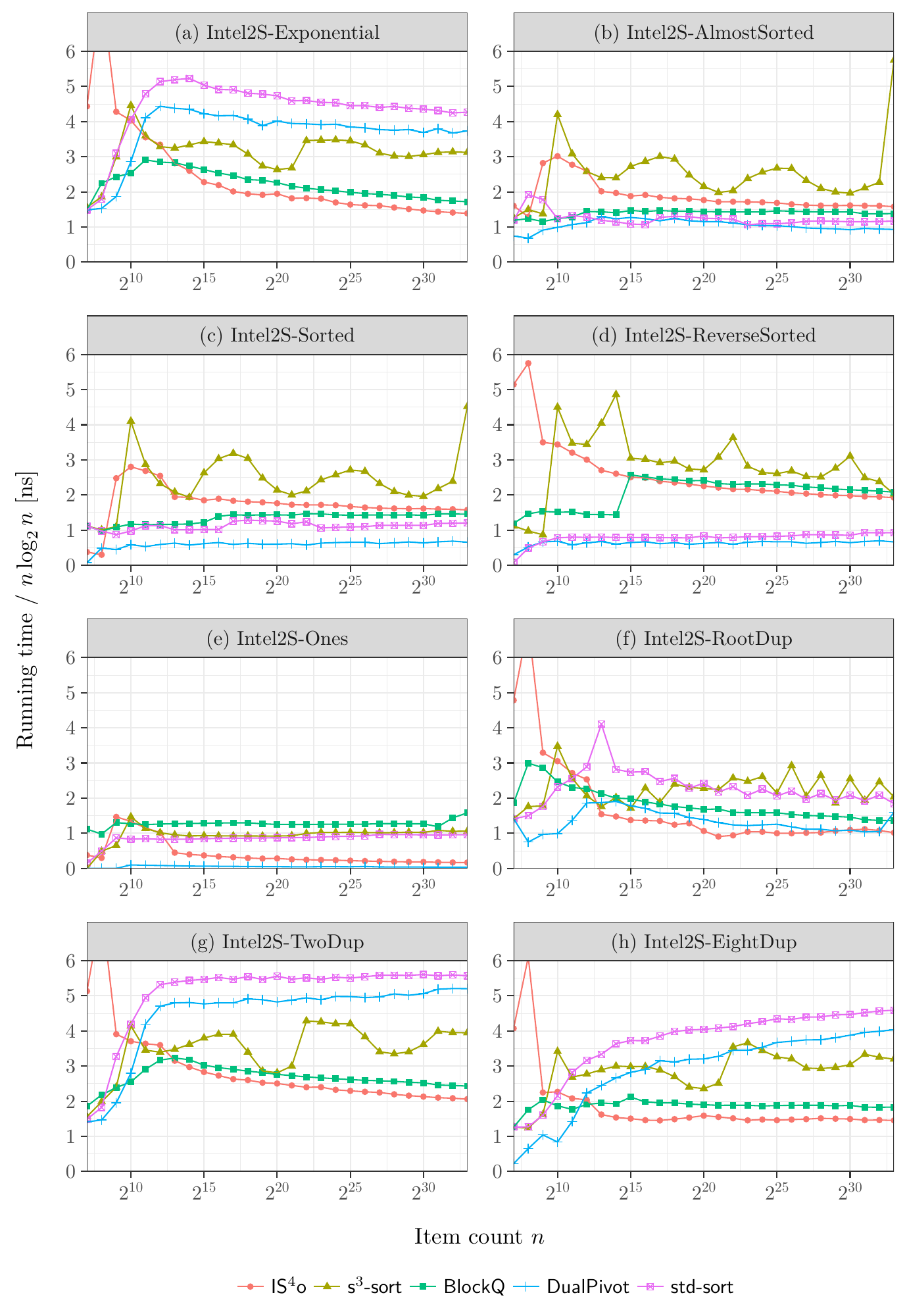}%
  \caption{\label{fig:sequential input distribution 132}%
    Running times of sequential algorithms on different input distributions executed on machine \pcinteltwo.}
\end{figure}
\begin{figure}[tbp]
  \centering%
  \includegraphics[]{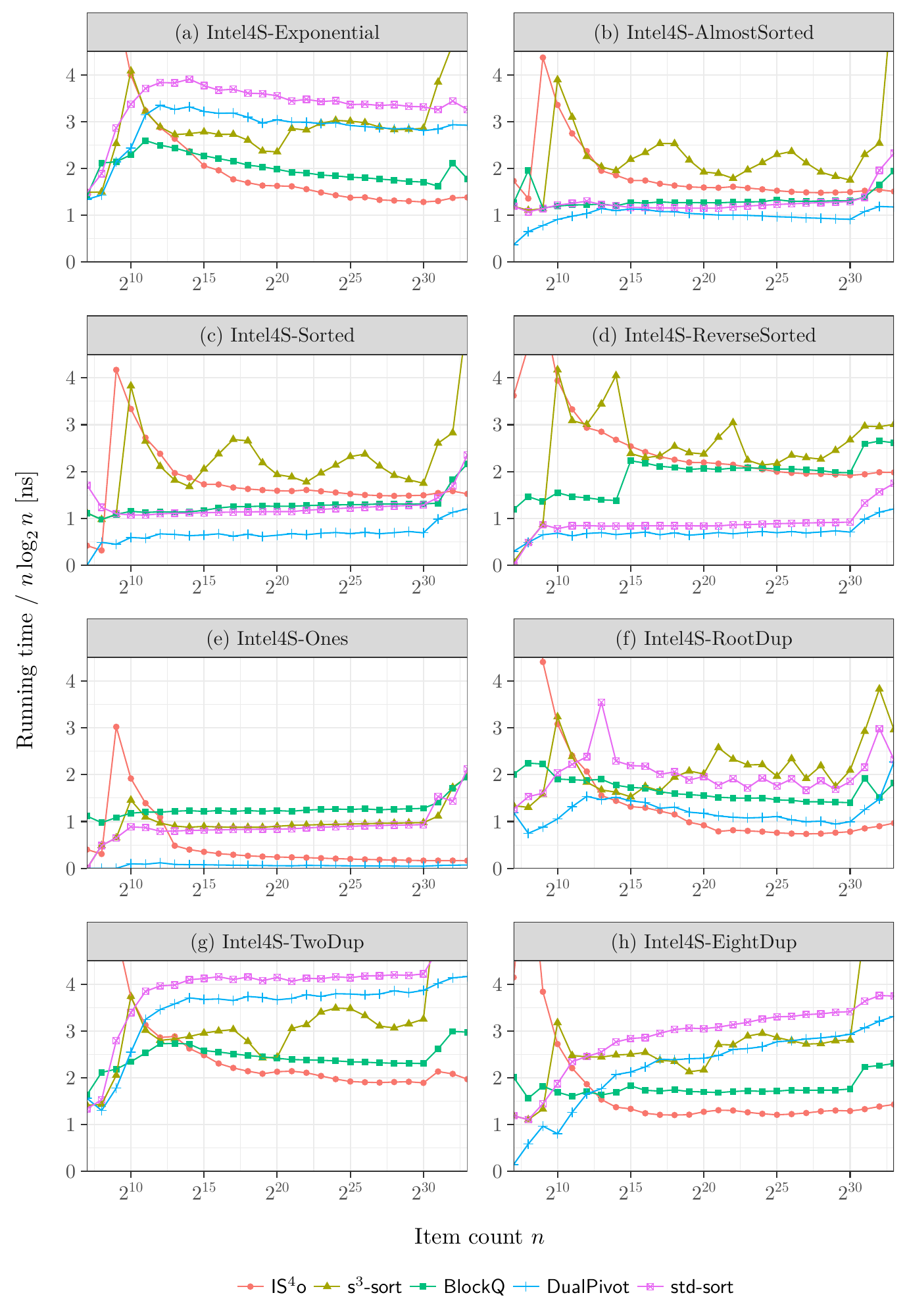}%
  \caption{\label{fig:sequential input distribution 127}%
    Running times of sequential algorithms on different input distributions executed on machine \pcintelfour.}
\end{figure}
\begin{figure}[tbp]
  \centering%
  \includegraphics[]{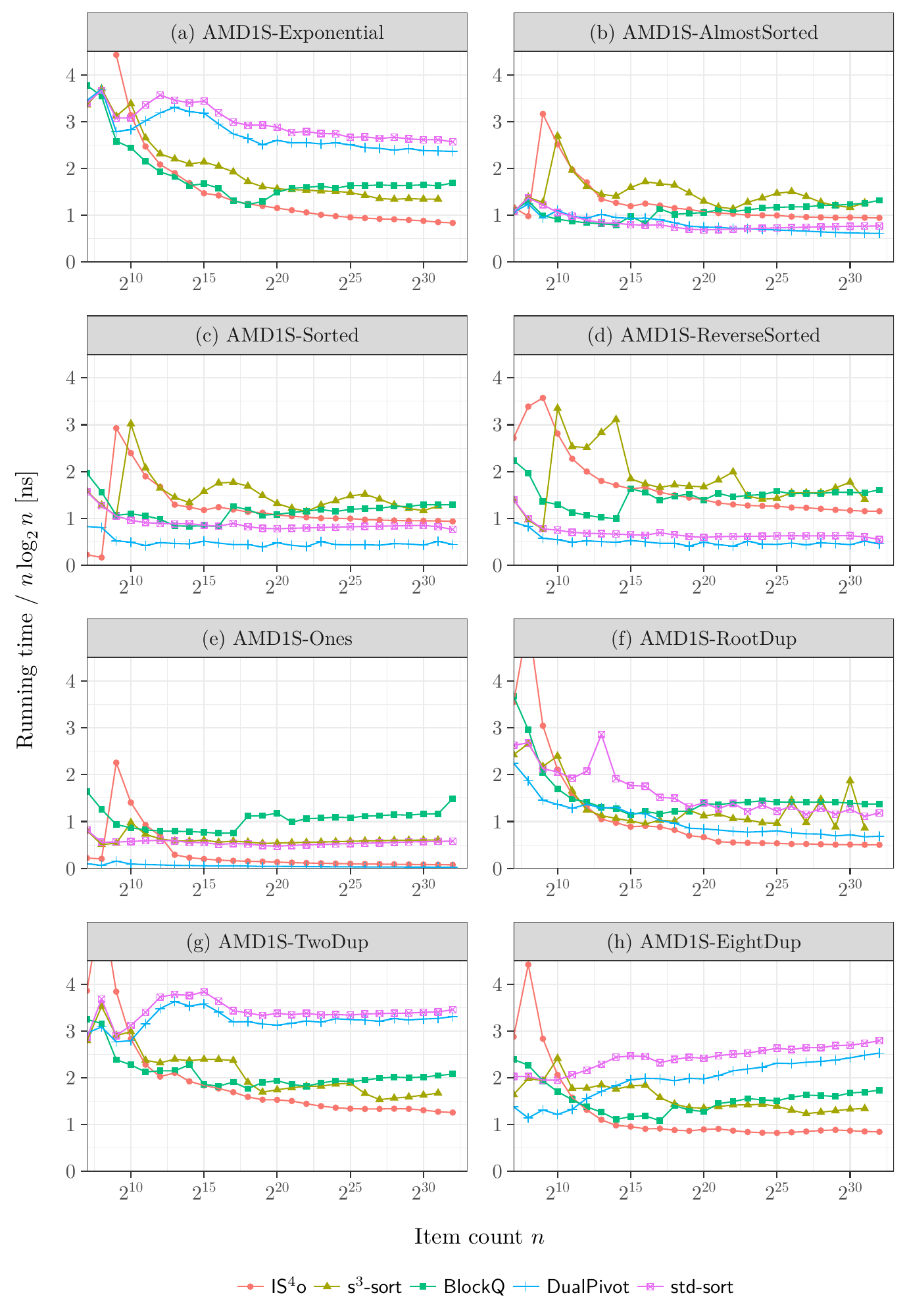}%
  \caption{\label{fig:sequential input distribution 133}%
    Running times of sequential algorithms on different input distributions executed on machine \pcamd.}
\end{figure}

\end{document}